\documentclass[Afour,sageh,times,doublespace]{sagej}

\usepackage{moreverb,url}
\usepackage{algorithm}
\usepackage{algorithmic}
\usepackage[shortlabels]{enumitem}
\usepackage[subrefformat=parens]{subcaption}

\usepackage[colorlinks,bookmarksopen,bookmarksnumbered,citecolor=red,urlcolor=red]{hyperref}

\newcommand\BibTeX{{\rmfamily B\kern-.05em \textsc{i\kern-.025em b}\kern-.08em
T\kern-.1667em\lower.7ex\hbox{E}\kern-.125emX}}

\def\volumeyear{2016}

\begin{document}

\runninghead{Ozaki, Uchino, and Imamura}

\title{Ozaki Scheme II:\\A GEMM-oriented emulation of\\floating-point matrix multiplication\\using an integer modular technique}

\author{Katsuhisa Ozaki\affilnum{1}, Yuki Uchino\affilnum{2}, and Toshiyuki Imamura\affilnum{2}}

\affiliation{\affilnum{1}Department of Mathematical Sciences, Shibaura Institute of Technology, Japan\\
\affilnum{2}RIKEN Center for Computational Science, Japan}

\corrauth{Katsuhisa Ozaki, 
	307 Fukasaku, Minuma-ku, Saitama-shi, Saitama 337-8570, Japan}

\email{ozaki@sic.shibaura-it.ac.jp}

\begin{abstract}
This paper addresses emulation algorithms for matrix multiplication. General Matrix-Matrix Multiplication (GEMM), a fundamental operation in the Basic Linear Algebra Subprograms (BLAS), is typically optimized for specific hardware architectures. The Ozaki scheme is a well-established GEMM-based emulation method for matrix multiplication, wherein input matrices are decomposed into several low-precision components to ensure that the resulting matrix product is computed exactly through numerical operations.
This study proposes a novel GEMM-based emulation method for matrix multiplication that leverages the Chinese Remainder Theorem.
The proposed method inherits the computational efficiency of highly optimized GEMM routines and further enables control over the number of matrix multiplications, which can enhance computational accuracy.
We present numerical experiments featuring INT8 Tensor Core operations on GPUs and FP64 arithmetic on CPUs as case studies. The results demonstrate that FP64 emulation using the proposed method achieves performance levels of up to 7.4 to 9.8 TFLOPS on the NVIDIA RTX 4090 and 56.6 to 80.2 TFLOPS on the NVIDIA GH200, exceeding the measured performance of native FP64 arithmetic. Furthermore, for FP64 computations on CPUs, the proposed method achieved up to a 2.3x speedup in emulating quadruple-precision arithmetic compared to the conventional Ozaki scheme.

\end{abstract}


\maketitle

\section{Introduction}
\label{sec:Introduction}

This paper discusses numerical methods for emulating high-precision matrix multiplication.  
When the accuracy of the computed results is insufficient, multiple-precision arithmetic provides a viable alternative~\citep{Higham2018}.  
For such computations, high-precision datatypes, such as the built-in float128\_t in the C++23 standard, and multiple-precision arithmetic libraries, such as MPFR~\citep{MPFR}, offer robust and reliable functionality.  
In the domain of linear algebra, MPLAPACK~\citep{Nakata2022MPLAPACK} is available for high-precision problem solving.
Alternatively, when it is unnecessary to extend the exponent range of floating-point numbers and only the significand requires pseudo-extension, \emph{multiple-word arithmetic} provides an efficient solution.
Such methods include double- and quad-word arithmetic~\citep{hida2001algorithms,QDLibrary}, as well as triple-word arithmetic~\citep{Muller_TripleWord}. 

When the computational task is limited to matrix multiplication, the Ozaki scheme~\citep{ozaki2012error,ozaki2013generalization} is recognized as a highly reliable method.  
It achieves high accuracy by leveraging standard floating-point operations and high performance by exploiting optimized routines, e.g.,  General Matrix-Matrix Multiplication (GEMM),  in Basic Linear Algebra Subprograms (BLAS).

In recent years, increasing attention has been paid to matrix engines optimized for low-precision arithmetic.  
From the perspective of power efficiency, mixed-precision computation utilizing low-precision formats has also garnered significant interest.  
Table~\ref{tab:Specifications} presents the floating-point and integer performance at various precisions on NVIDIA GPUs.
Note that the specification of FP16 Tensor Cores (TCs) on RTX 4090 (165.2 TFLOPS) is for FP16 input and FP32 output, the specifications for the H100 and GH200 are the same as those for the H200.
Tensor Cores are specialized matrix multiply--accumulate units in GPUs provided by~\cite{tensorcore}.
While FP64 performance shows a substantial gap between data-center-class and consumer-grade GPUs, FP16 and INT8 TCs operations offer outstanding throughput.

\begin{table}[htb]
\centering
\caption{Specifications in TFLOPS/TOPS of GPUs~\citep{tensorcore} for dense data}
\label{tab:Specifications}
\begin{tabular}{@{}l@{ }l@{\ }c@{\ }c@{\ }c@{\ }c@{}}
    \hline
    &     & Ampere & ADA & Hopper & Blackwell \\ 
    &    & A100 SXM4 & RTX 4090 & H200 SXM5 & B200\\
    &    & GA100 & AD102-300 & & \\ \hline
    FP64 &    & 9.7  &  1.29   &    34     &  40 \\
    FP64 & TC & 19.5 &    --    &    67     &  40 \\
    FP32 &    & 19.5 &  82.6   &    67     &  80 \\
    TF32 & TC & 156  &  82.6   &    494    & 1100 \\
    BF16 & TC & 312  &   165.2   &    989    & 2250 \\
    FP16 & TC & 312  & 165.2 &    989    & 2250 \\
    INT8 & TC & 624  &   660.6   &   1979    & 4500 \\
    FP8  & TC &  --   &   660.6   &   1979    & 4500 \\
    FP6  & TC &    --    &  --   &     --     & 4500 \\
    INT4 & TC &  --   &  1321.2   &     --     &  --  \\
    FP4  & TC &    --   &  --   &     --     & 9000  \\ \hline
\end{tabular}
\end{table}

The use of FP16 TCs in the Ozaki scheme has been discussed~\citep{mukunoki2020}. 
Subsequently, \cite{ootomo2024dgemm} employed the Ozaki scheme with INT8 TCs to emulate matrix multiplication in FP64, and \cite{ozIMMU} released the API \texttt{ozIMMU} as open source~\cite{ozIMMUGit}.
\cite{Uchino2025} further accelerated \texttt{ozIMMU}, and the enhanced version is also publicly released~\citep{ozIMMU-uchino}.  
The potential applications of the Ozaki scheme are discussed~\citep{Dawson2024}. 
In addition, Dongarra et al.~\citep{Dongarra2024} reported preliminary HPL 
performance and power-efficiency results on an NVIDIA Blackwell B200 GPU. 
In their HPL experiment, the DGEMM-dominated trailing matrix updates arising 
in the LU factorization for solving dense linear systems were evaluated using 
FP64 emulation based on the Ozaki scheme with 7 slices, and were compared with 
native FP64 computation.

In this study, we propose a novel matrix multiplication emulation method based on the Chinese Remainder Theorem (CRT).
CRT has long played a significant role in exact computation, and the related framework is often referred to as the Residue Number System (RNS).
Exact integer arithmetic for dot products based on CRT has been studied since the early days of electronic computing; see, e.g., \cite{takahashi1960new}.
Accurate inner products for rational inputs have also been studied using modular arithmetic techniques by~\cite{emiris1998modular}.
RNS- and CRT-based approaches have been investigated in several other studies, e.g., \cite{Isupov2020EfficientGPU,papachristou1988systolic,doliskani2018simultaneous}.
There are also libraries for modular arithmetic and related exact computations: NTL~\citep{ShoupNTL}, FLINT~\citep{flint}, RNS-APAL~\citep{OlsenRNSAPALSoftware}, Mathemagix~\citep{van_der_Hoeven_Lecerf_Quintin_2016}, and FFLAS-FFPACK~\citep{DumasGiorgiPernet2008FFLASFFPACK}.
Some of these studies and libraries use CRT-based arithmetic, BLAS routines, or both.
To the best of our knowledge, however, there appears to be no existing CRT-based approach that combines all three aspects considered here: floating-point matrix inputs, CRT-based computation, and the use of BLAS routines.

Similarly to emulation methods based on the conventional Ozaki scheme, the proposed approach is also GEMM-based, allowing the use of INT8 matrix engines.
In contrast to the conventional Ozaki scheme, the proposed method significantly reduces the number of required matrix multiplications, which is a notable advantage.
While the conventional Ozaki scheme controls accuracy by adjusting the number of slices, the proposed method enables precise control over the number of matrix multiplications.

We present numerical results obtained using INT8 TCs on the NVIDIA GH200 Grace Hopper Superchip and the NVIDIA GeForce RTX 4090 GPU, as well as FP64 operations on the Intel\textsuperscript{\textregistered} Core\texttrademark{} i7-8665U processor and the Intel\textsuperscript{\textregistered} Core\textsuperscript{\texttrademark} i9-10980XE processor.
The proposed method achieved 56.6--80.2 TFLOPS in FP64-equivalent precision on the GH200, and 7.4--9.8 TFLOPS on the RTX 4090.  
Given that the measured performance of FP64 TCs is 61.9 TFLOPS, the emulation achieves a performance that is comparable to or even exceeds native FP64.  
In addition, for quad-word arithmetic emulation on a CPU, the proposed method achieved an approximate 2.3x speedup compared to the conventional Ozaki scheme.

The remainder of this paper is organized as follows.  
Section~\ref{sec:Notation and Previous Study} introduces the notation, the conventional Ozaki scheme, and the Chinese Remainder Theorem.  
Section~\ref{sec:Proposed Method} describes the proposed method (referred to as Ozaki scheme~II) and outlines algorithms using INT8 TCs on GPU and FP64 arithmetic on CPU.  
Section~\ref{sec:Numerical Experiment} presents numerical experiments conducted on GPUs and CPUs to evaluate the effectiveness of the proposed method.  
Finally, Section~\ref{sec:Conclusion} concludes the paper.

\section{Notation and Previous Study}\label{sec:Notation and Previous Study}

\subsection{Notation}

Let $\mathbb{F}$ be a set of binary floating-point numbers as defined by~\cite{ieee754}.
We define a constant $u$ as the unit roundoff, e.g., 
$u=2^{-53}$ for FP64.
Let $\mathbb{Z}_k$ for $k \in \mathbb{N}$ be a set of integers, where $a \in \mathbb{Z}_k$ means $|a| \le 2^{k}$.
The notation $\mathtt{fl}(\cdot)$ indicates a computed result using floating-point arithmetic.
We assume that neither overflow nor underflow occurs in $\mathtt{fl}(\cdot)$.
For a matrix $A = (a_{ij})$, the notation $|A|$ represents the matrix whose entry $(i,j)$ is $|a_{ij}|$; that is, $|A| = (|a_{ij}|)$.
The notation $\gcd(a, b)$ stands for the greatest common divisor of two integers $a$ and $b$.

The modulo operation $a \bmod m$, where $a \in \mathbb{Z}$ and $m \in \mathbb{N}$, can be defined in various ways. One such variant is the symmetric modulo, in which the remainder $r$ satisfies 
\[
-\frac{m}{2} \le r \le \frac{m}{2}.
\]
In this paper, we adopt this symmetric definition.
Formally, $r = a \bmod m$ is defined by
\begin{equation}\label{def:mod}
r = a - m \cdot \left\lfloor \frac{a}{m} + \frac{1}{2} \right\rfloor.
\end{equation}
This definition ensures that the remainder is the integer closest to zero among those congruent to $a$ modulo $m$. 
For a matrix $A$, the expression $A \bmod m$ refers to applying the modulo operation to each element of the matrix. 
This results in a new matrix of the same dimensions as $A$.

\subsection{Conventional Ozaki scheme}

For $A \in \mathbb{F}^{p \times q}$ and $B \in \mathbb{F}^{q \times r}$, our aim is to obtain an approximation of $AB$.
The Ozaki scheme~\citep{ozaki2012error,ozaki2013generalization} splits the matrices into
\begin{equation}
    \begin{aligned}
        A &= A_1 + A_2 + \dots + A_{k-1} + \underline{A}_k, \\
        B &= B_1 + B_2 + \dots + B_{\ell-1} + \underline{B}_\ell, 
    \end{aligned}
    \label{eq:spliting}
\end{equation}
where
\[
A_i \in \mathbb{F}^{p \times q}, \quad B_j \in \mathbb{F}^{q \times r}
\]
for $1 \le i \le k-1$ and $1 \le j \le \ell-1$,  
and 
\[
\underline{A}_k := A - \sum_{i=1}^{k-1} A_i \in \mathbb{F}^{p \times q}, \quad 
\underline{B}_{j} := B - \sum_{i=1}^{j-1} B_i \in \mathbb{F}^{q \times r}
\]
for $1 \le j \le \ell$.
Here, $\underline{A}_k$ denotes the remainder after approximating $A$ by the first $k-1$ slices, and 
$\underline{B}_\ell$ is defined analogously.
For \eqref{eq:spliting}, 
we call $k$ the number of slices for $A$ and $\ell$ that for $B$.
Here, we set $k = \ell$ according to~\cite{ozaki2012error}, but it is also possible to set $k \not= \ell$ as discussed by~\cite{ozaki2013generalization}.
Then, $AB$ is transformed into  
\begin{equation}
AB = \sum_{i+j \le k} A_i B_j + \sum_{i=1}^{k-1} A_i \underline{B}_{k+1-i} + \underline{A}_k B.
\label{eq:OzIform}    
\end{equation}
The matrix products $A_i B_j$ for $i + j \le k$ can be computed without rounding errors using floating-point arithmetic.
Note that \cite{Ozaki2025} also proposed an alternative form of $AB$, but this is not considered in this paper.

The Ozaki scheme for $k = \ell$ consists of the following three parts:
\begin{description}
    \item[Part 1:] splitting matrices as in~\eqref{eq:spliting}
    \item[Part 2:] computation of $k(k+1)/2$ matrix products as in~\eqref{eq:OzIform}
    \item[Part 3:] reduction of the computed matrix products
\end{description}
The computational costs are $\mathcal{O}(pq)+\mathcal{O}(qr)$ for Part 1, $k(k+1)pqr$ for Part 2, and $\mathcal{O}(pr)$ for Part 3.
Thus, Part 2 dominates the overall cost for sufficiently large $p$, $q$, and $r$.
One advantage of the Ozaki scheme is that optimized BLAS routines can be applied to this computationally intensive part.
However, a disadvantage is that a large amount of memory is required to store the matrices, as they are decomposed into multiple summands in~\eqref{eq:spliting}.
In Part 2, an appropriate BLAS routine is selected based on the structure of the matrices.
For example, if $A$ is a triangular matrix and $B$ is a general matrix, TRMM is used.
If $A$ and $B$ have no special structure, GEMM is used. 

Even when $A$ and $B$ are represented as multi-word formats, a similar approach can achieve high-precision computational results.
For example, let $A \in \mathbb{R}^{p \times q}$ and $B \in \mathbb{R}^{q \times r}$ be represented as double-word formats: $A := A_h + A_\ell$ and $B := B_h + B_\ell$ for $A_h, A_\ell \in \mathbb{F}^{p \times q}$ and $B_h, B_\ell \in \mathbb{F}^{q \times r}$ such that
\[
\mathtt{fl}(A_h + A_\ell) = A_h, \quad 
\mathtt{fl}(B_h + B_\ell) = B_h.
\]
Similarly, we divide $A_h + A_\ell$ and $B_h + B_\ell$ into 
the unevaluated sum of floating-point matrices such that
\begin{equation}
    \begin{aligned}
        A_h + A_\ell &\approx A_1 + A_2 + \dots + A_{e-1} + \underline{A}_e, \\
        B_h + B_\ell &\approx B_1 + B_2 + \dots + B_{f-1} +  \underline{B}_f,
    \end{aligned}
    \label{eq:spliting2}
\end{equation}
where $A_i \in \mathbb{F}^{p \times q}$, $B_j \in \mathbb{F}^{q \times r}$, and $\mathtt{fl}(A_i B_j) = A_i B_j$ for $1 \le i \le e-1$ and $1 \le j \le f-1$.
In addition, 
\begin{align*}
\mathbb{F}^{p \times q} & \ni \underline{A}_e \approx A_h + A_\ell - \sum_{i=1}^{e-1} A_i, \\
\mathbb{F}^{q \times r} & \ni
\underline{B}_f \approx B_h + B_\ell - \sum_{i=1}^{f -1}B_i.
\end{align*}
Then, we obtain a computed result based on~\eqref{eq:OzIform}.
Note that in the calculations of Part 3, high-precision computations such as double-word arithmetic are required in this case.

Next, we introduce a method for emulating matrix multiplication using matrix engines available in recent GPUs. 
Typical examples include NVIDIA TCs.
Let nonsingular diagonal matrices $D_i \in \mathbb{F}^{p \times p}$ and $E_i \in \mathbb{F}^{r \times r}$ whose diagonal elements are powers of two.
This aim is to achieve an error-free diagonal scaling.
\cite{mukunoki2020} set 
\begin{equation}
    \begin{aligned}
        A & \approx D_1^{-1} D_1 A_1 + D_2^{-1} D_2 A_2 + \dots + D_k^{-1} D_k A_k, \\
        B & \approx B_1 E_1 E_1^{-1} + B_2 E_2 E_2^{-1} + \dots + B_k E_k E_k^{-1}, 
    \end{aligned}
\label{eq:Mukunoki}
\end{equation}
where all elements in $D_i A_i$ and $B_i E_i$ for all $1 \le i \le k$ can be represented in FP16.
The entries of the products $D_i A_i$ and $B_i E_i$ behave as though they possess a significand with
\begin{equation}
\left\lfloor \frac{24 - \log_2 q}{2} \right\rfloor
\label{eq:mukunoki_bit}    
\end{equation}
bits of precision.
The entries of \(D_i A_i\) and \(B_i E_i\) are treated as quantities with at most $\lfloor (24 - \log_2 q)/2 \rfloor$
effective significand bits. This choice ensures that the product of two such
quantities and the subsequent accumulation of \(q\) terms fit within the 24-bit significand of FP32 arithmetic.
Then, $(D_i A_i)(B_j E_j)$ can be computed without rounding error using FP16 TCs.
Note that FP16 TCs can store the results in either FP16 or FP32, but in this case, the computed results are stored in FP32.
Then, the approximation $\tilde C$ is obtained by
\[
\tilde C \approx \sum_{i + j \le k + 1} D_i^{-1} ((D_i A_i) (B_j E_j)) E_j^{-1}.
\]
This involves $k(k+1)/2$ matrix multiplications.
After computing matrix multiplication $(D_i A_i) (B_j E_j)$ using FP16 TCs, the results are converted from FP32 to FP64, and a summation and scaling of these is performed in FP64.

\cite{ootomo2024dgemm} similarly used the form~\eqref{eq:Mukunoki} where all elements of $D_i A_i$ and $B_i E_i$ for all $i$ are stored in INT8.
Taking into account that the results of INT8 TCs operations are stored in INT32, they utilized this property to achieve error-free computation of the product $(D_i A_i) (B_j E_j)$ when $q \le 2^{29}$ using INT8 TCs.
The first advantage of using INT8 TCs is that it is theoretically twice or 4x as fast as FP16 TCs as in Table~\ref{tab:Specifications}.
The second advantage of using INT8 TCs is that if $q < 2^{17}$, the significand of the split matrices is kept 7-bit independent of $q$.
If $q>2^{10}$, the bits~\eqref{eq:mukunoki_bit} for FP16 TCs is less than seven.

The acceleration of ozIMMU and the analysis of rounding errors are discussed by~\cite{Uchino2025}.
They achieved acceleration by reducing the cost of the reduction in Part~3. 
Specifically, they reduced the number of summation calculations performed using FP64.
In this paper, we call this type of algorithm \emph{Ozaki scheme I}.
Ozaki scheme I using INT8 TCs consists of the following three parts:
\begin{description}
    \item[Part 1:] splitting matrices as in~\eqref{eq:Mukunoki}
    \item[Part 2:] computation of $k(k+1)/2$ matrix products using INT8 TCs
    \item[Part 3:] reduction of the computed matrix products in FP64.
\end{description}

\begin{figure*}[htb]\centering
\noindent
\begin{minipage}[b]{0.49\hsize}\centering
\includegraphics[width=\hsize]{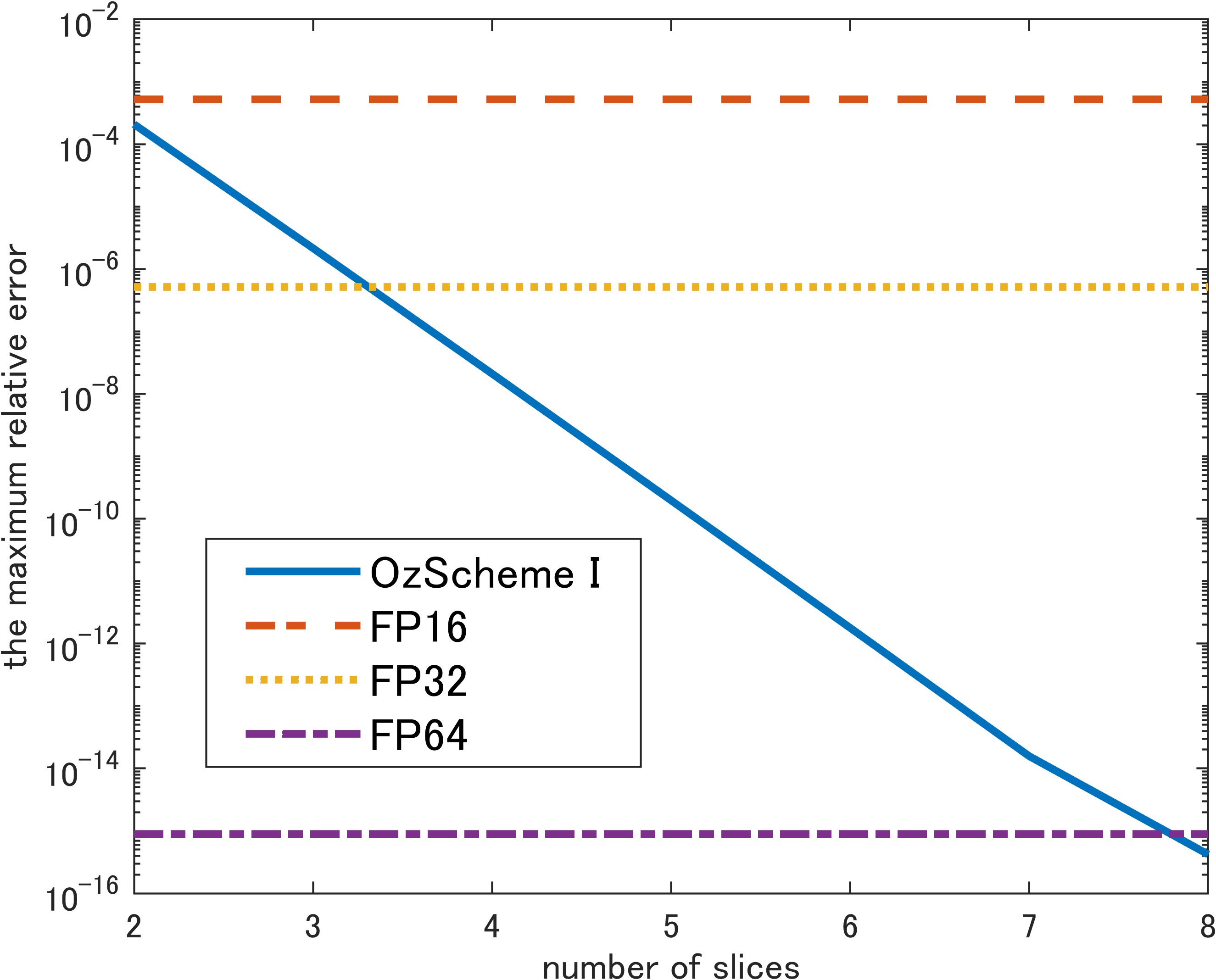}
\end{minipage}
\begin{minipage}[b]{0.49\hsize}\centering
\includegraphics[width=\hsize]{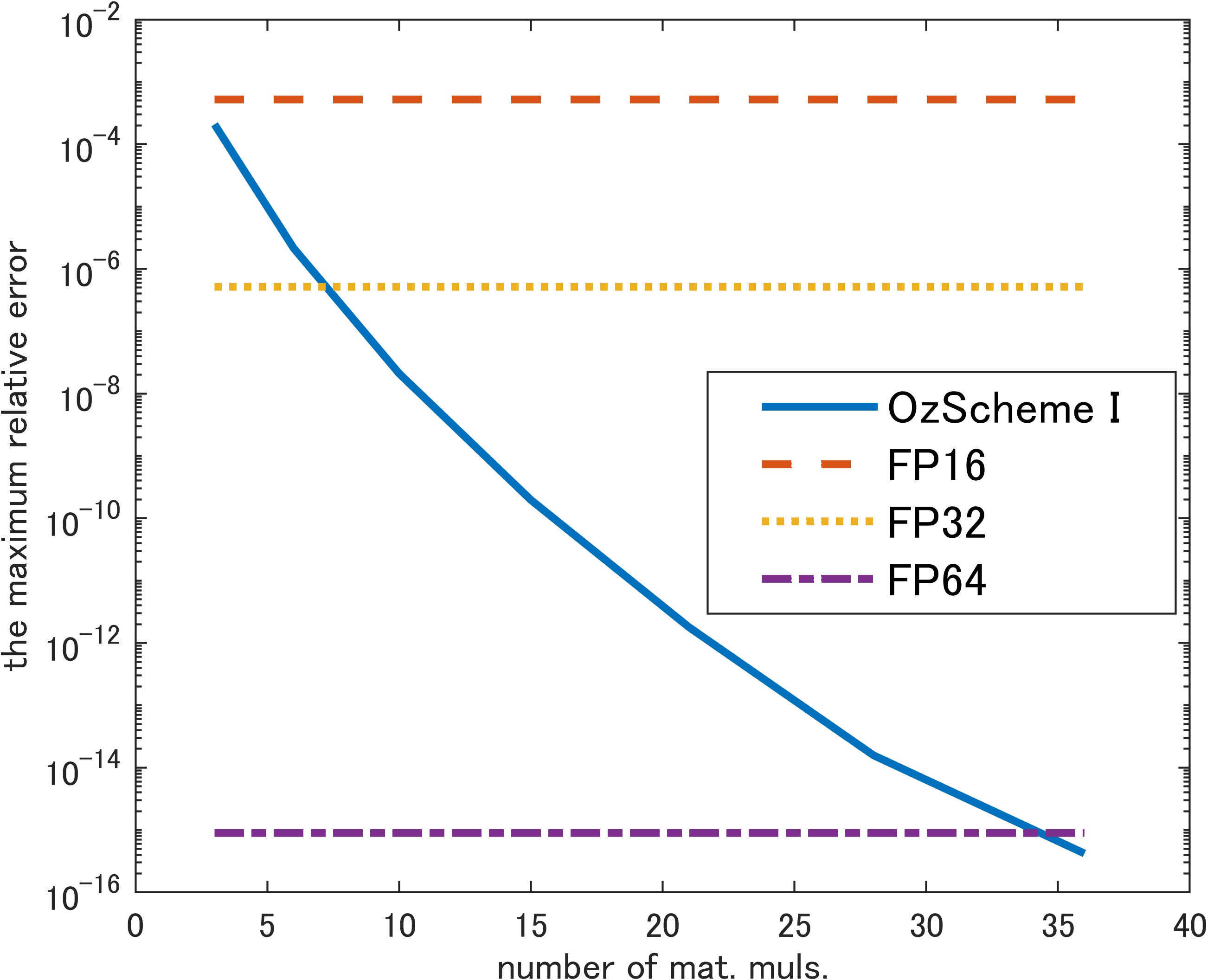}
\end{minipage}

\caption{Maximum relative error of AB for FP16, FP32, FP64, and OZ-I with various numbers of slices (left), and the number of matrix multiplications (right)}
\label{fig:Oz1slices_mul}
\end{figure*}

Here, we introduce the accuracy trends of Ozaki scheme I using INT8 TCs.
In this case, $D_i A_i$ and $B_i E_i$ for all $i$ in \eqref{eq:Mukunoki} are represented by INT8.
We generated two matrices using MATLAB as
\begin{equation}
A = \mathtt{randn}(1000), \quad B = \mathtt{randn}(1000), 
\label{eq:testmat}
\end{equation}
where $\mathtt{randn}(n)$ generates an 
$n \times n$ matrix of normally distributed random numbers with mean 0 and variance 1.
Table~\ref{tab:slices} shows the number of matrix multiplications required for each slice.
Figure~\ref{fig:Oz1slices_mul} shows the maximum relative error of the result computed by Ozaki scheme I
for~\eqref{eq:testmat} for the number of slices (left) and the matrix multiplications (right).
From Fig.~\ref{fig:Oz1slices_mul}, Ozaki scheme I improves in accuracy in proportion to the number of slices.
When increasing the number of slices from $k$ to $k+1$, 
$k+1$ additional matrix multiplications are required. 
Therefore, as shown in Fig.~\ref{fig:Oz1slices_mul}, the accuracy does not exhibit linear growth with respect to the number of matrix multiplications when plotted on a logarithmic scale.

\begin{table}[htb]
    \centering
    \caption{Relation between the number of slices and the number of matrix multiplications}
    \label{tab:results}
    \begin{tabular}{c|ccccccccc} 
        \toprule
        slices & 2 & 3 & 4 & 5 & 6 & 7 & 8 & 9 & 10\\
        \midrule
        muls. & 3 & 6 & 10 & 15 & 21 & 28 & 36 & 45 & 55\\
        \bottomrule
    \end{tabular}
    \label{tab:slices}
\end{table}

\subsection{Chinese Remainder Theorem}

A brief introduction to the Chinese Remainder Theorem is provided, as it is essential for the proposed method.
Let $m_1, m_2, \dots, m_s$ be pairwise coprime positive integers, i.e., $\gcd(m_i, m_j) = 1$ for all $i \neq j$. 
Let \(a_1,a_2,\dots,a_s\) be integers, and define
\begin{equation}
M := \prod_{i=1}^s m_i .
\label{eq:defM}
\end{equation}
Then the system of congruences
\[
\begin{aligned}
    x &\equiv a_1 \pmod{m_1}, \\
    x &\equiv a_2 \pmod{m_2}, \\
    &\vdots \\
    x &\equiv a_s \pmod{m_s}
\end{aligned}
\]
has a unique solution modulo \(M\).
The solution can be explicitly constructed as
\[
x \equiv \sum_{i=1}^s a_i M_i y_i \bmod{M},
\]
where $M_i = M / m_i$ and $y_i$ are the modular multiplicative inverses of $M_i$ modulo $m_i$, that is, $M_i y_i \equiv 1 \bmod{m_i}$.

We apply this technique to matrix multiplication.
Let $A' \in \mathbb{Z}^{p \times q}$ and $B' \in \mathbb{Z}^{q \times r}$.
Let
\[
C_i \equiv A' B' \bmod {m_i}.
\]
Then, we have
\begin{align}
C & \equiv A'B' \bmod M \nonumber \\
& = \sum_{i=1}^s C_i M_i y_i \bmod M.
\label{eq:C}
\end{align}
In this paper, we use this approach, a direct reconstruction using Chinese Remainder Theorem.

\section{Proposed Method}\label{sec:Proposed Method}

In this section, we propose a GEMM-based method that combines the Chinese Remainder Theorem and error-free matrix multiplication, which we refer to as \emph{Ozaki scheme~II}.
Let $A \in \mathbb{F}^{p \times q}$ and $B \in \mathbb{F}^{q \times r}$.
We prepare two diagonal matrices $D$ and $E$ whose diagonal elements are powers of two, and we have
\begin{equation}
C = AB = D^{-1} D AB E E^{-1} \approx D^{-1} A' B' E^{-1}, 
\label{eq:scaling}    
\end{equation}
where 
\begin{equation}
DA \approx A' \in \mathbb{Z}^{p \times q}_{k_A}, \quad 
BE \approx B' \in \mathbb{Z}^{q \times r}_{k_B}.
\label{eq:def_k}
\end{equation}
The diagonal elements of $D$ and $E$ are
\begin{align*}
d_{ii} & = 2^{k_A-1} \cdot \max_{1 \le j \le q, \ a_{ij}\not=0} \mathtt{ufp}\left( \frac{1}{|a_{ij}|}\right), \\
e_{jj} & = 2^{k_B-1} \cdot \max_{1 \le i \le q, b_{ij} \not = 0} \mathtt{ufp}\left( \frac{1}{|b_{ij}|}\right),
\end{align*}
where $\mathtt{ufp}(\cdot)$ indicates the unit in the first place.

We first set $s$, the number of matrix multiplications.
We next pick up $m_i \ge 2$ for $1 \le i \le s$ from a table,
where $m_1, m_2, \dots, m_s$ are coprime to each other, 
and set $M$ as~\eqref{eq:defM}.
Matrices $A'_t \in \mathbb{Z}^{p \times q}$ and $B'_t \in \mathbb{Z}^{q \times r}$ are generated as follows:
\begin{equation}
    \begin{aligned}
(a'_t)_{ij} & \equiv a'_{ij} \bmod {m_t}, \quad -\frac{m_t}{2} \le (a'_t)_{ij} \le \frac{m_t}{2}, \\
(b'_t)_{ij} & \equiv b'_{ij} \bmod {m_t}, \quad -\frac{m_t}{2} \le (b'_t)_{ij} \le \frac{m_t}{2}
    \end{aligned}
\label{eq:modAB}
\end{equation}
for $1 \le t \le s$.
If the moduli $m_i$ are chosen so that all inner products appearing in $A'_t B'_t$ are exactly representable in the working precision, then $A'_t B'_t$ can be computed without rounding errors using a BLAS routine.
This condition is discussed in detail in Sections~3.1 and~3.2, using INT8 TCs and FP64 as examples.

\begin{figure}[htbp]
    \centering
    \includegraphics[width=\hsize]{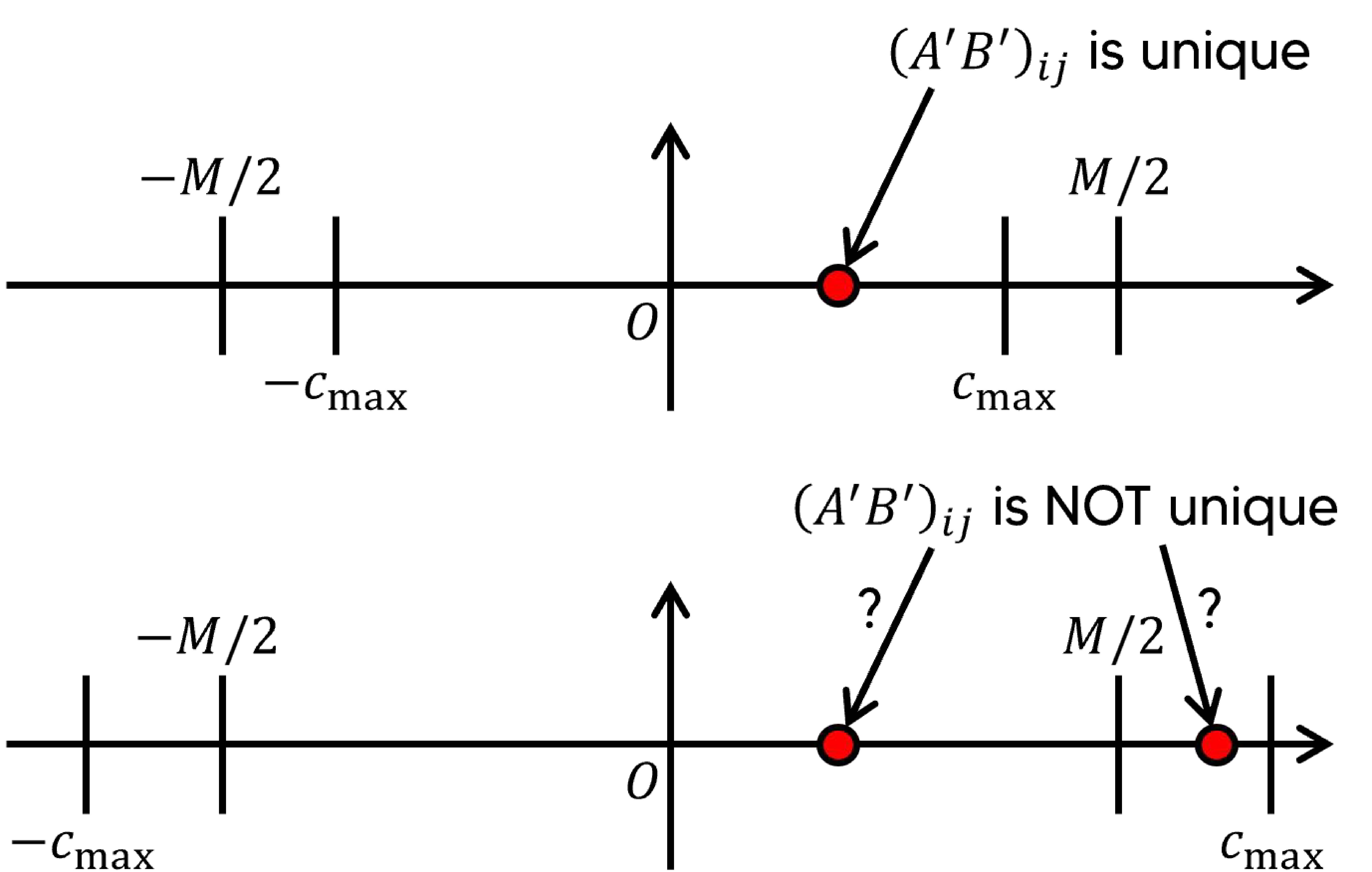} 
    \caption{Unique / Non-unique candidate for $A'B'$} 
    \label{fig:range} 
\end{figure}

Let $C_i \equiv A'_i B'_i \bmod {m_i}$ and $Y = \sum_{i=1}^s C_i M_i y_i$.
Then, we have
\begin{equation}
C' \equiv A'B' \bmod M = Y \bmod M.
\label{eq:crt}
\end{equation}
Therefore, the candidates of $c'_{ij}$ are  
\begin{align*}
\dots, \ y_{ij} - 2M, \ y_{ij} - M, \ y_{ij}, \ y_{ij} + M, \ y_{ij} + 2M, \dots \ .
\end{align*}
Assume that for all $(i,j)$ pairs 
\[
c_{\min} \le (A'B')_{ij} \le c_{\max}.
\]
If $c_{\max} - c_{\min} < M$, we find the unique $C'$ in~\eqref{eq:crt}.
For simplicity, let
\[
c_{\max} = -c_{\min}:= \max_{i,j} (|A'||B'|)_{ij}.
\]
If 
\begin{equation}
2 c_{\max} < M
\label{eq:unique_condition}   
\end{equation}
is satisfied, we can find the matrix $X \in \mathbb{Z}^{p \times r}$ such that
\begin{equation}
-\frac{M}{2} < - c_{\max} \le X_{ij} \le c_{\max} < \frac{M}{2}.
\label{eq:rangeX}   
\end{equation}
If $M$ is smaller than or equal to $2 c_{\max}$, we may find multiple candidates of the result 
(see Fig.~\ref{fig:range}).
Although $c_{\max}$ is defined as the maximum over all elements for simplicity, an element-wise definition is also possible.

The constants $k_A$ and $k_B$ in \eqref{eq:def_k} are important for the accuracy of the computed result because it is strongly related to the truncation error.
Here we explain a simple way to find $k_A$ and $k_B$.
From~\eqref{eq:def_k}, 
\begin{equation}
\max_{i,j} \left( |A'| |B'| \right)_{ij} \le 
q 2^{k_A + k_B} < \frac{M}{2}.
\label{eq:upper_bound}
\end{equation}
From~\eqref{eq:upper_bound}, we set $k_A + k_B$ in~\eqref{eq:def_k} as
\begin{equation}
k_A + k_B := \left\lfloor \log_2 \frac{M/2 - 1}{q} \right\rfloor.
\label{eq:easy_k}
\end{equation}
Then, we can set the diagonal matrices $D$ and $E$ in~\eqref{eq:def_k}.
If $A$ and $B$ have the same precision, we assume $k_A = k_B$, leading to the following expression:
\begin{equation}
k_A = k_B := \left\lfloor \frac{1}{2}\log_2 \frac{M/2 - 1}{q} \right\rfloor.
\label{eq:easy_k2}
\end{equation}
If the precisions of the matrices $A$ and $B$ differ, for instance, if $A$ is in FP32 and $B$ is in FP64, we assume $k_A < k_B$.
If $s$ increases, $M$ also increases, and as a result, 
$k_A + k_B$ becomes larger.
Note that 
\begin{itemize}
    \item if the matrices are sparse, $q$ in~\eqref{eq:upper_bound} can be reduced,
    \item $q 2^{k_A + k_B}$ in~\eqref{eq:upper_bound} is overestimated as the upper bound of $c_{\max}$. Alternative ways are to use the Cauchy-Schwarz inequality or to use low-precision computation (see Section 3.1), 
    \item if the range of $A'B'$ is known, that is, $c_{\min}$ and $c_{\max}$, we can improve~\eqref{eq:easy_k}. 
\end{itemize}

Overall, Ozaki scheme II consists of the following four parts:
\begin{description}
    \item[Part 1:] computation of $k_A$ and $k_B$ from the given $s$, $A$, and $B$. Then obtain $A'$ and $B'$ as in~\eqref{eq:def_k}.
    \item[Part 2:] repetition of the following for $i=1, \dots, s$
    \begin{description}
        \item[Part 2-a:] determination of $A'_i$ and $B'_i$ as in~\eqref{eq:modAB}.  
        \item[Part 2-b:] computation of $C_i:=A'_iB'_i$ using a BLAS routine.
        \item[Part 2-c:] computation of $Z:= Z + C_i M_i y_i$.
    \end{description}
    \item[Part 3:] determination of the unique candidate $X$ from the matrix $Z$.
    \item[Part 4:] application of the inverse scaling $D^{-1} X E^{-1}$ in \eqref{eq:scaling}.
\end{description}

Note that $M_i y_i$ and $m_i$ are calculated in advance and stored in a table.
In Part 2-b, we use appropriate functions such as GEMM, TRMM, or SYRK depending on the structure of the matrices $A$ and $B$.
In Part 3, we find $x_{ij}$ in~\eqref{eq:rangeX} by a range reduction, so the required working precision is higher than the precision of the final result in Part 2-c.
In Ozaki scheme~I, we keep $A_1, \dots, A_k$ and $B_k, \dots, B_\ell$, which consume a lot of memory.
However, Ozaki scheme II immediately discards the matrices $A_i'$ and $B_i'$ after obtaining $C_i$.

\subsection{Using INT8 TCs for matrix multiplication}

If we use INT8 TCs on GPU, we set $2 \le m_i \le 256$ for $1 \le i \le s$ and
$m_1, m_2, \dots, m_s$ to be coprime to each other.
Under these conditions, we prepare the largest possible $m_1, m_2, \dots, m_s$.
For example, when $s=16$, we set $m_i$ as
\begin{equation}
    \begin{split}
        m &:= (256, \ 255, \ 253, \ 251, \\
        &\qquad 247, \ 239, \ 233, \ 229, \\
        &\qquad 227, \ 223, \ 217, \ 211, \\ 
        &\qquad 199, \ 197, \ 193, \ 191)^T \in \mathbb{N}^{16}.
    \end{split}
    \label{eq:int8_law8}
\end{equation}
These numbers and $M_i y_i$ in~\eqref{eq:C} are stored in a table for $s = 2,3,\dots\ $.
The result of modular arithmetic with any number modulo $m_i$ falls within the range $-128$ to $127$, which is represented in INT8.
In the case where the result of a modulo 256 operation is 128, the wraparound behavior of the INT8 type maps this value to $-128$ and $128 \equiv -128 \bmod 256$, thereby avoiding any issues.
Then no error occurs in $A'_i B'_i$ for $q < 2^{17}$ using
cublasGemmEx because the result is stored by INT32.
If $q \ge 2^{17}$, it is possible to apply block matrix multiplication for error-free matrix multiplication.
The code has been open-sourced by~\cite{GEMMul8}.

\begin{figure}[htb]\centering
\noindent
\includegraphics[width=\hsize]{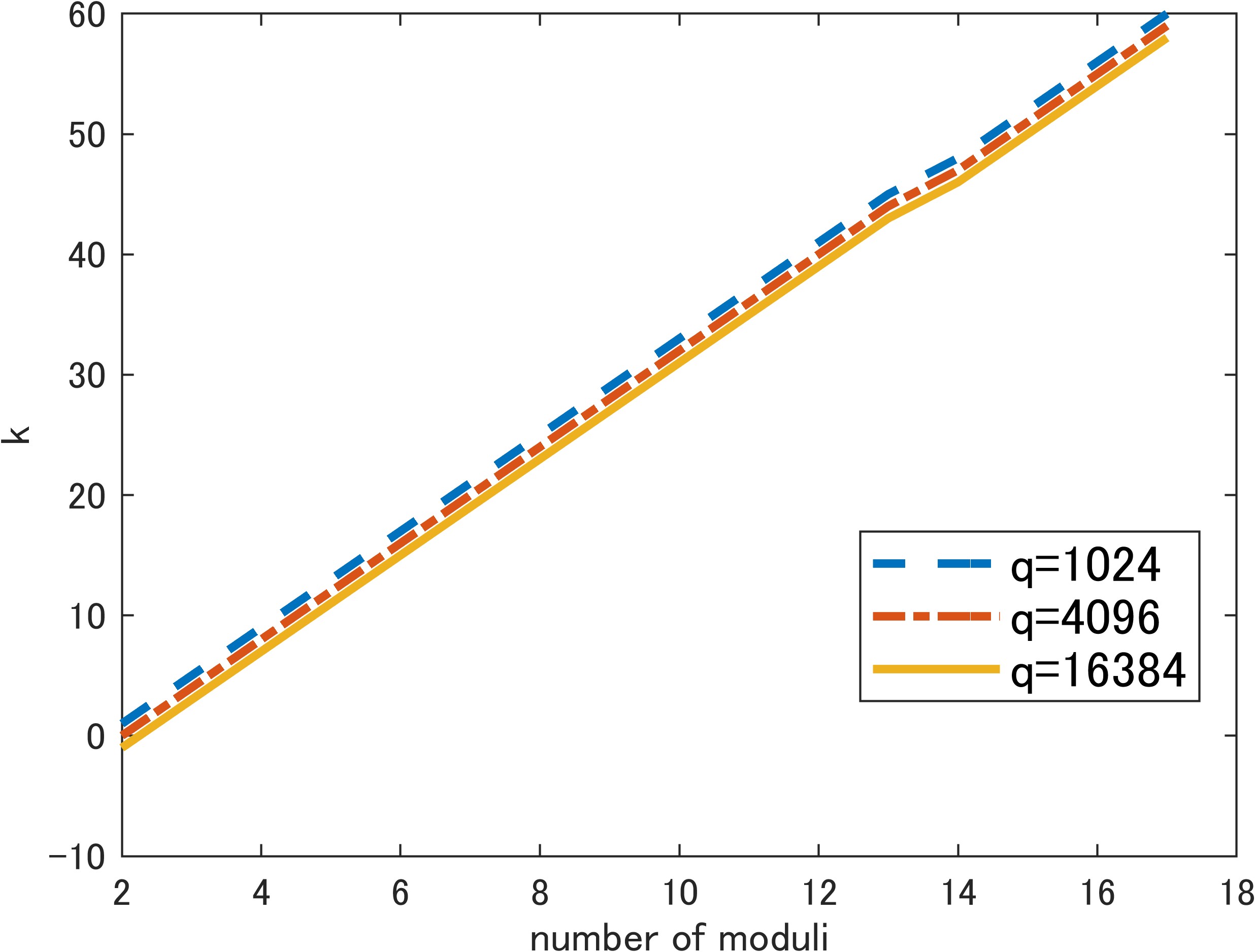}
\caption{$k:=k_A = k_B$ in~\eqref{eq:easy_k2} from number of moduli using INT8 Tensor Cores for matrix multiplication}
\label{fig:estimate-int8}
\end{figure}

Figure~\ref{fig:estimate-int8} shows $k:= k_A = k_B$ in~\eqref{eq:easy_k2} for $p = q = r \in \{1024, 4096, 16384\}$.
For $s=16$, $k = 53$ is expected to be achieved, allowing FP64 emulation with 16 matrix multiplications.
Here, this expectation is true if there are no significant differences in the absolute values between the elements of the matrix.
To achieve results comparable to double-precision arithmetic, Ozaki scheme I requires 7 to 8 slices, corresponding to 28 to 35 matrix multiplications. This observation suggests the potential advantage of Ozaki scheme II.
Note that this should be understood as an expectation based on the underlying fixed-point approximation; the actually achieved accuracy depends on the scaling of the input matrices and on the distribution of the magnitudes of their entries.

\begin{algorithm}[htb]
\caption{%
    Outline of proposed method using INT8 tensor cores. %
    The functions $\mathrm{trunc}(\cdot)$ and $\mathrm{round}(\cdot)$ round inputs into integers in round-towards-zero and round-to-nearest-even mode.
    $\mathbb{F}_{64}$ is a set of floating-point numbers in FP64. %
    The labels (F), (I), and (H) indicate operations performed using 
    FP64 arithmetic,
    integer arithmetic, and 
    high-precision arithmetic,
    respectively.%
    \label{alg:Oz2-int8}%
}
\begin{algorithmic}[1]
\REQUIRE $A \in \mathbb{F}_{64}^{p \times q}$, $B \in \mathbb{F}_{64}^{q \times r}$, $m \in \mathbb{N}^s$, $s \in \mathbb{N}$

\item[\textbf{Convert FP64 to INT8:}]{\small
    \STATE\label{alg:Oz2-int8-1} (F) Determine shift values $D \in \mathbb{F}_{64}^{p \times q}$ and $E \in \mathbb{F}_{64}^{q \times r}$
    \STATE\label{alg:Oz2-int8-2} (F) $A' := \mathrm{trunc}(DA)$
    \hfill\COMMENT{$A' \in \mathbb{F}_{64}^{p \times q} \cap \mathbb{Z}^{p \times q}$}
    \STATE\label{alg:Oz2-int8-3} (F) $B' := \mathrm{trunc}(BE)$
    \hfill\COMMENT{$B' \in \mathbb{F}_{64}^{q \times r} \cap \mathbb{Z}^{q \times r}$}
    \STATE (I) $(a'_t)_{ij} := a'_{ij} \bmod m_t$ $(1 \le t \le s)$
    \STATE (I) $(b'_t)_{ij} := b'_{ij} \bmod m_t$ $(1 \le t \le s)$

\item[\textbf{Matrix multiplication using INT8 TCs:}]
    \STATE (I) $C'_t := A'_tB'_t$ $(1 \le t \le s)$
    
\item[\textbf{Convert matrix products into UINT8:}]
    \STATE\label{alg:Oz2-int8-7} (I) $(c''_t)_{ij} := (c'_t)_{ij} - \lfloor (c'_t)_{ij} / m_t \rfloor m_t$ $(1 \le t \le s)$

\item[\textbf{Accumulate matrix products and inversely scale:}]
    \STATE (H) Compute $M$ in~\eqref{eq:defM} in advance.
    \STATE\label{alg:Oz2-int8-8} (H) $C''' := \sum_{t=1}^s C''_t\cdot My_t/m_t$
    \STATE\label{alg:Oz2-int8-9} (H) $C''' := C''' \bmod M$
    \STATE\label{alg:Oz2-int8-10} (F) $C := D^{-1}C'''E^{-1}$

\ENSURE $C \in \mathbb{F}_{64}^{p \times r}$ 
}\end{algorithmic}
\end{algorithm}

\subsection{Using FP64 for matrix multiplication}

We set prime numbers $m_1, m_2, \dots, m_s$ as
\begin{equation}
q m_i^2 \le 4 u^{-1}, \quad 1 \le i \le s.
\label{eq:double_bound}
\end{equation}
For $A'_t \in \mathbb{Z}^{p \times q}$ and $B'_t \in \mathbb{Z}^{q \times r}$ in \eqref{eq:modAB}, using~\eqref{eq:double_bound}, we have
\begin{equation}
(|A'_t| |B'_t|)_{ij} \le q \cdot \frac{1}{2}m_t \cdot \frac{1}{2}m_t \le u^{-1}, \quad u = 2^{-53}
\label{eq:FP64up}
\end{equation}
for all $(i,j)$ pairs and $1 \le t \le s$.
Hence, no rounding error occurs in $A'_i B'_i$ for $1 \le i \le s$ using GEMM in BLAS.
Similarly, considering~\eqref{eq:upper_bound}, $k_A + k_B$ is obtained as in~\eqref{eq:easy_k}.

Note that $m_i$ for $1\le i \le s$ does not need to be a prime number as long as they are pairwise coprime. 
However, unlike the case of INT8 TCs, when using FP64, they are simply chosen to be prime numbers to ensure that sufficiently large primes can be easily found.
For example, for $s=16$ and $n = 2^{10}$, we set $m_i$ from~\eqref{eq:double_bound} as
\begin{equation}\label{eq:fp64_law8}
    \begin{split}
        m &:= (4194301, 4194287, 4194277, 4194271, \\
        &\qquad 4194247, 4194217, 4194199, 4194191, \\
        &\qquad 4194187, 4194181, 4194173, 4194167, \\ 
        &\qquad 4194143, 4194137, 4194131, 4194107)^T \in \mathbb{N}^{16}.
    \end{split}
\end{equation}
The reason is that from~\eqref{eq:double_bound}, 
\[
m_i^2 \le \frac{2^{55}}{2^{10}}, 
\]
so that
\[
m_1 \approx 2^{22} \le \sqrt{2^{45}}.
\]
Note that $m$ in \eqref{eq:fp64_law8} satisfies
\[
\frac{m_{16}}{m_1} = \frac{4194107}{4194301} = 0.99995\dots\ .
\]
It is expected that
\[
M \approx s m_s.
\]
Because $ M $ is proportional to $s$, $k_A + k_B$ is also proportional to $s$.
In contrast, $m$ in \eqref{eq:int8_law8} satisfies
\[
\frac{m_{16}}{m_1} = \frac{191}{256} = 0.74\dots \ .
\]
The growth rate of $ M $ decreases as $ s $ increases.

Here, we deal with the application of FP64 to multi-word arithmetic.
Ozaki scheme II can be applied to any multi-word format; we assume that matrices are represented as an unevaluated sum of $v$ floating-point matrices such as
\begin{align}
A := \sum_{i=1}^v A^{(i)}, \quad B := \sum_{i=1}^v B^{(i)}, 
\label{eq:multi-word}
\end{align}
where for $1 \le i \le v-1$, 
\begin{align}
u |A^{(i)}| \ge |A^{(i+1)}|, \quad u |B^{(i)}| \ge |B^{(i+1)}|.
\label{eq:diff}
\end{align}
Again, we set diagonal matrices $D$ and $E$ such that
\[
C \approx D^{-1} A' B' E^{-1}, \quad A' \approx DA, \quad B' \approx BE.
\]
In Part 2-a, we compute
\begin{align*}
A_i' & \approx \left( D \sum_{i=1}^v A^{(i)} \right) \bmod {m_i}, \\
B_i' & \approx \left( \left( \sum_{i=1}^v B^{(i)} \right) E \right) \bmod {m_i},
\end{align*}
where $A'_i \in \mathbb{Z}^{p \times q}$ and $B'_i \in \mathbb{Z}^{q \times r}$ for all $i$.
The rest of the parts are the same as in the original.

\begin{figure}[htb]\centering
\includegraphics[width=\hsize]{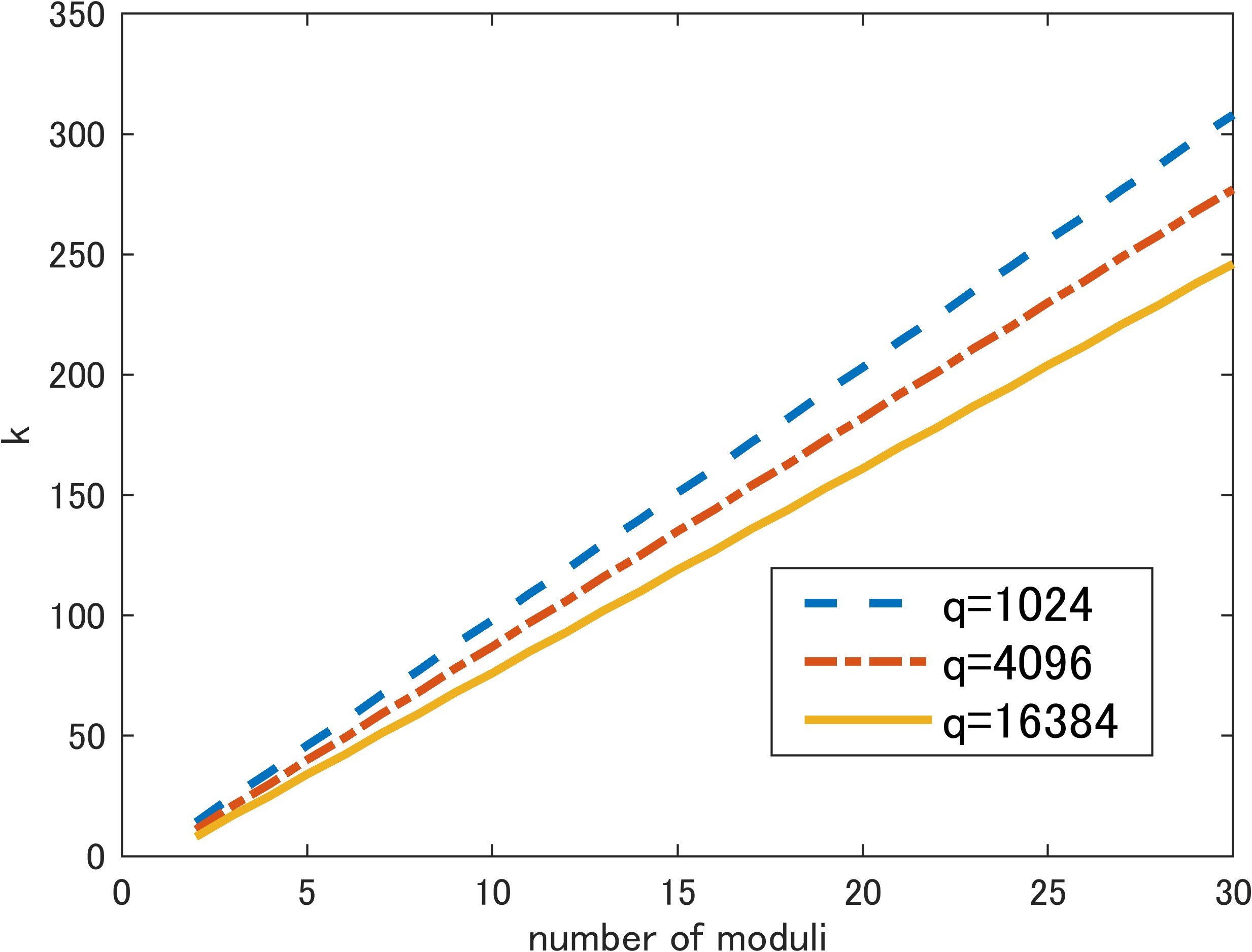}
\caption{$k:=k_A = k_B$ from number of moduli~\eqref{eq:easy_k2} using FP64 for matrix multiplication}
\label{fig:estimate}
\end{figure}

In the following discussion, we assume that the absolute values of the entries of the input matrices do not vary significantly.
Figure~\ref{fig:estimate} illustrates the behavior of the constant~$k := k_A = k_B$ as a function of the number of moduli~$s$. 
When $s$ is in the range of 16 to 20, the value of $k$ is approximately 160, which corresponds to a precision level equivalent to that obtained by triple-word arithmetic.  
As $s$ increases to the range of 21 to 25, $k$ increases to approximately 210, indicating a precision level comparable to that of quadruple-word arithmetic.

Thus far, we have considered the product of two matrices; however, the same approach can be extended to the product of three or more matrices.
As an example, when dealing with the product of three matrices $A$, $B$, and $C$, we convert them to
\begin{align*}
DA & \approx A' \in \mathbb{Z}^{p \times q}_{k_A}, \quad 
BE \approx B' \in \mathbb{Z}^{q \times r}_{k_B}, \\
E^{-1}CF & \approx C' \in \mathbb{Z}^{r \times w}_{k_C}, 
\end{align*}
where the matrices $D$, $E$, and $F$ are diagonal, with diagonal entries that are powers of two, and $k_A$, $k_B$, and $k_C$ are determined to satisfy 
\[
qr 2^{k_A + k_B + k_C} < \frac{M}{2}.
\]
Then a similar procedure to that in \eqref{eq:upper_bound} and \eqref{eq:easy_k} can be applied.
If $k_A = k_B = k_C$, we set
\[
k_A = k_B = k_C := \left\lfloor \frac{1}{3}\log_2 \frac{M/2 - 1}{qr} \right\rfloor.
\]

Note that the Ozaki-scheme-based emulation considered in this paper is intended to reproduce binary64-like accuracy for ordinary finite matrix products. It is not intended as a complete bit-level implementation of all IEEE 754 binary64 semantics. Thus, exceptional values such as NaNs, infinities, and signed zeros, whether present in the inputs or produced by the corresponding binary64 operation, require separate handling, for example, by detection and fallback to a native binary64 routine.

\section{Numerical Experiment}\label{sec:Numerical Experiment}

In this section, we present considerations on both accuracy and computational performance based on the results of numerical experiments. 
We report experimental results on FP64 emulation using INT8 TCs, as well as on quad-word format emulation using FP64 on a CPU.
In both scenarios, we assume that the precision levels are equal, that is, $k_A = k_B$.

\subsection{Using INT8 TCs for FP64 emulation}
All numerical experiments here were conducted on NVIDIA GH200 Grace Hopper Superchip and NVIDIA GeForce RTX 4090 GPU with NVIDIA CUDA Toolkit 12.8.61.
The tested methods will be denoted as follows:
\begin{itemize}
    \item DGEMM: cublasGemmEx in FP64
    \item OS II-fast-$s$:  Algorithm~\ref{alg:Oz2-int8} with $s$ moduli, employing the Cauchy--Schwarz inequality for the line~\ref{alg:Oz2-int8-1} to satisfy the condition~\eqref{eq:unique_condition} 
    \item OS II-accu-$s$:  Algorithm~\ref{alg:Oz2-int8} with $s$ moduli, employing INT8 matrix multiplication for the line~\ref{alg:Oz2-int8-1} to satisfy the condition~\eqref{eq:unique_condition} 
\end{itemize}
To satisfy condition~\eqref{eq:unique_condition}, the diagonal matrices
$D$ and $E$ are chosen so that
\[
    \left( |A'| |B'| \right)_{ij}
    \le D_{ii} \left( |A| |B| \right)_{ij} E_{jj}
    < M/2 .
\]
The two variants OS II-fast-$s$ and OS II-accu-$s$ differ in how they estimate an upper bound for $\left( |A| |B| \right)_{ij}$.
In the fast variant, this quantity is bounded using the Cauchy--Schwarz inequality:
\[
    \left( |A| |B| \right)_{ij} \le \|A(i,:)\|_2 \|B(:,j)\|_2 .
\]
The scaling factors $D_{ii}$ and $E_{jj}$ are then determined in reverse so that
\[
    D_{ii} \|A(i,:)\|_2 \|B(:,j)\|_2 E_{jj} < M/2 .
\]
In the accu variant, $|A|$ and $|B|$ are first bounded by INT8 matrices $A''$ and $B''$ as
\[
    |A| \le D'' A'',\quad |B| \le B'' E'',
\]
where $D''$ and $E''$ are diagonal scaling matrices. 
This gives
\[
    \left( |A| |B| \right)_{ij}
    \le D''_{ii} \left( A''B'' \right)_{ij} E''_{jj}
    =: C''_{ij}.
\]
Here, the product $A''B''$ is computed using INT8 TCs. 
The scaling factors $D_{ii}$ and $E_{jj}$ are then determined in reverse so that
\[
    D_{ii} C''_{ij} E_{jj} < M/2 .
\]

The test matrices $A \in \mathbb{F}^{p \times q}$ and $B \in \mathbb{F}^{q \times r}$ were generated as
\[
    a_{ij},b_{ij} \approx  (\mathrm{rand}-0.5)\cdot \exp(\phi\cdot \mathrm{randn}),
\]
where $\phi \in \mathbb{R}$ controls the exponent distribution, 
$\mathrm{rand} \in (0,1] \subset \mathbb{R}$ is a uniformly distributed random number and
$\mathrm{randn} \in \mathbb{R}$ is drawn from the standard normal distribution.
The non-negative constant $\phi$ specifies the tendency to difficulty in terms of the accuracy of matrix multiplication.

\begin{figure*}[htb]\centering
\includegraphics[width=\hsize]{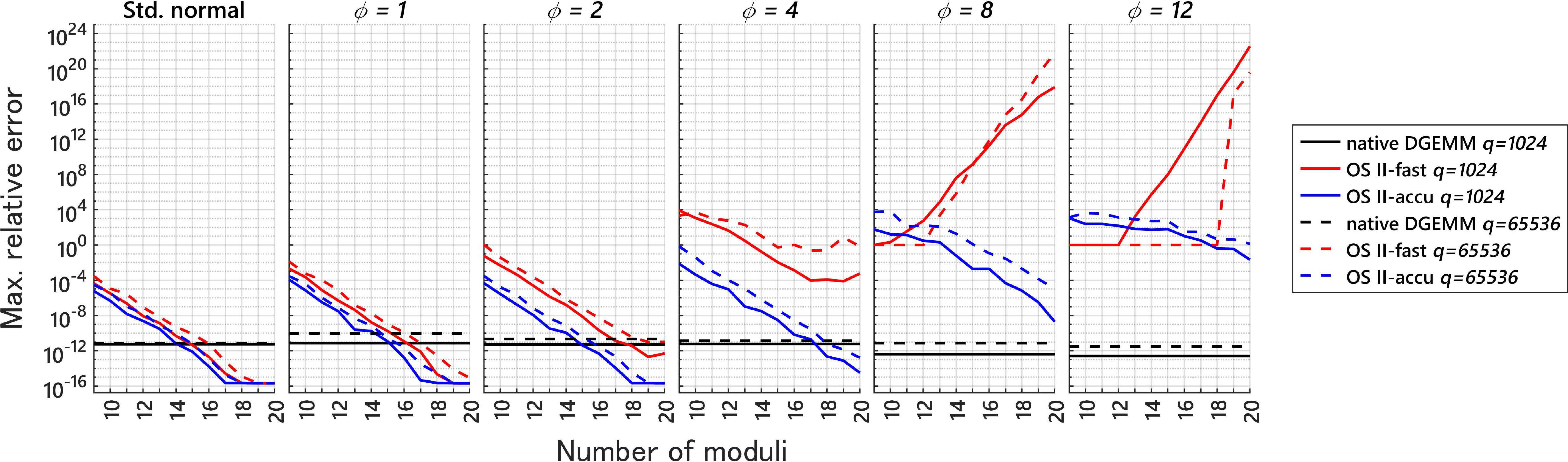}
\caption{Accuracy comparison for $p = r = 1024$ and $q \in \{1024, 2048, 4096, 8192, 16384\}$ on NVIDIA GH200 Grace Hopper Superchip}
\label{fig:int8-accuracy}
\end{figure*}

Figure~\ref{fig:int8-accuracy} shows the accuracy of DGEMM, OS II-fast-$s$, and OS II-accu-$s$ for $s \in \{8,9,\dots,20\}$ on GH200.
We obtained similar results on RTX 4090.
The initial estimation at the line~\ref{alg:Oz2-int8-1} of Algorithm~\ref{alg:Oz2-int8} strongly affects the truncation error at the lines~\ref{alg:Oz2-int8-2} and \ref{alg:Oz2-int8-3}.
Thus, OS II-accu returns more accurate results than OS II-fast due to less overestimation of the upper bound of $|A'||B'|$ in~\eqref{eq:unique_condition} by direct matrix multiplication using INT8 TCs.
The limiting accuracy of OS II-fast and OS II-accu depends on the arithmetic precision of lines~\ref{alg:Oz2-int8-8} and~\ref{alg:Oz2-int8-9} in Algorithm~\ref{alg:Oz2-int8}. In the implementation, pseudo-Double-Double arithmetic was used.
For $\phi = 0.5$, both proposed methods require $14$ or $15$ moduli to achieve the accuracy of the DGEMM level.
OS II-accu can deal with larger $\phi$ to produce sufficiently accurate results.

Tables~\ref{tab:int8-performance-gh200} and \ref{tab:int8-performance-rtx} show the throughput performance of DGEMM, 
OS II-fast-$s$, and OS II-accu-$s$ for $s \in \{14,15,16,17,18\}$.
For small problems, matrix multiplication using INT8 TCs did not achieve sufficient performance. 
Additionally, the overhead of operations except for matrix multiplication became relatively significant, resulting in substantially lower performance compared to DGEMM.
However, as shown in Table~\ref{tab:int8-performance-gh200}, our implementation achieved $80.2$ TFLOPS for $p=q=r=16384$ and $s = 14$ on GH200.
These results demonstrate that the proposed methods achieved higher throughput than DGEMM while attaining DGEMM-level accuracy.

\begin{table}[htb]
\centering
\caption{Throughput in TFLOPS on NVIDIA GH200 Grace Hopper Superchip\label{tab:int8-performance-gh200}}
\begin{tabular}{@{}lcccc@{}}
\hline
    Methods $\backslash$ Matrix size & $2048$ & $4096$ & $8192$ & $16384$  \\
\hline
    DGEMM (cuBLAS)    & 56.9   & 61.3   & 62.0   & 60.9  \\
    OS II-fast-$14$   & 16.5   & 48.6   & 72.1   & 80.2  \\
    OS II-fast-$15$   & 15.4   & 45.2   & 67.4   & 74.6  \\
    OS II-fast-$16$   & 14.6   & 43.2   & 63.8   & 70.4  \\
    OS II-fast-$17$   & 13.8   & 40.5   & 60.1   & 66.1  \\
    OS II-fast-$18$   & 13.0   & 38.3   & 56.9   & 62.6  \\
    OS II-accu-$14$   & 14.7   & 42.7   & 64.7   & 71.1  \\
    OS II-accu-$15$   & 14.0   & 40.2   & 60.9   & 66.9  \\
    OS II-accu-$16$   & 13.3   & 38.5   & 58.0   & 63.3  \\
    OS II-accu-$17$   & 12.6   & 36.4   & 55.1   & 59.9  \\
    OS II-accu-$18$   & 12.0   & 34.7   & 52.2   & 56.6  \\
\hline
\end{tabular}
\end{table}

\begin{table}[htb]
\centering
\caption{Throughput in TFLOPS on NVIDIA RTX 4090\label{tab:int8-performance-rtx}}
\begin{tabular}{@{}lccc@{}}
\hline
    Methods $\backslash$ Matrix size & $2048$ & $4096$ & $8192$  \\
\hline
    DGEMM (cuBLAS)    & 0.61   & 0.62   & 0.62  \\
    OS II-fast-$14$   & 3.73   & 6.94   & 9.81  \\
    OS II-fast-$15$   & 3.54   & 6.55   & 9.20  \\
    OS II-fast-$16$   & 3.31   & 6.22   & 8.76  \\
    OS II-fast-$17$   & 3.07   & 5.90   & 8.29  \\
    OS II-fast-$18$   & 2.91   & 5.56   & 7.83  \\
    OS II-accu-$14$   & 3.46   & 6.51   & 9.23  \\
    OS II-accu-$15$   & 3.29   & 6.04   & 8.68  \\
    OS II-accu-$16$   & 3.10   & 5.69   & 8.26  \\
    OS II-accu-$17$   & 2.93   & 5.40   & 7.81  \\
    OS II-accu-$18$   & 2.84   & 5.18   & 7.41  \\
\hline
\end{tabular}
\end{table}

\begin{figure*}[htb]\centering
\noindent
\begin{minipage}[b]{0.49\hsize}\centering
\includegraphics[width=\hsize]{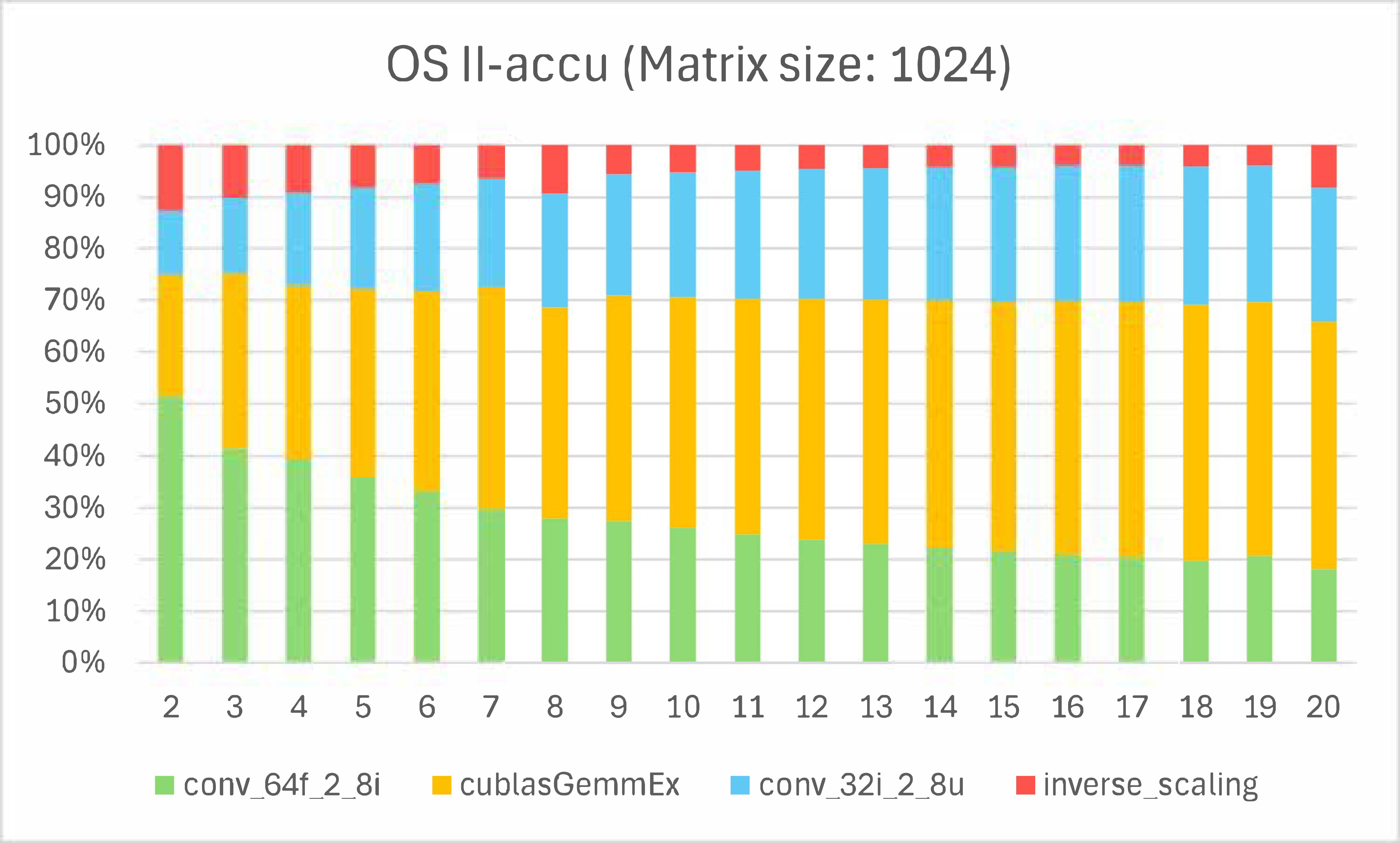}
\end{minipage}
\begin{minipage}[b]{0.49\hsize}\centering
\includegraphics[width=\hsize]{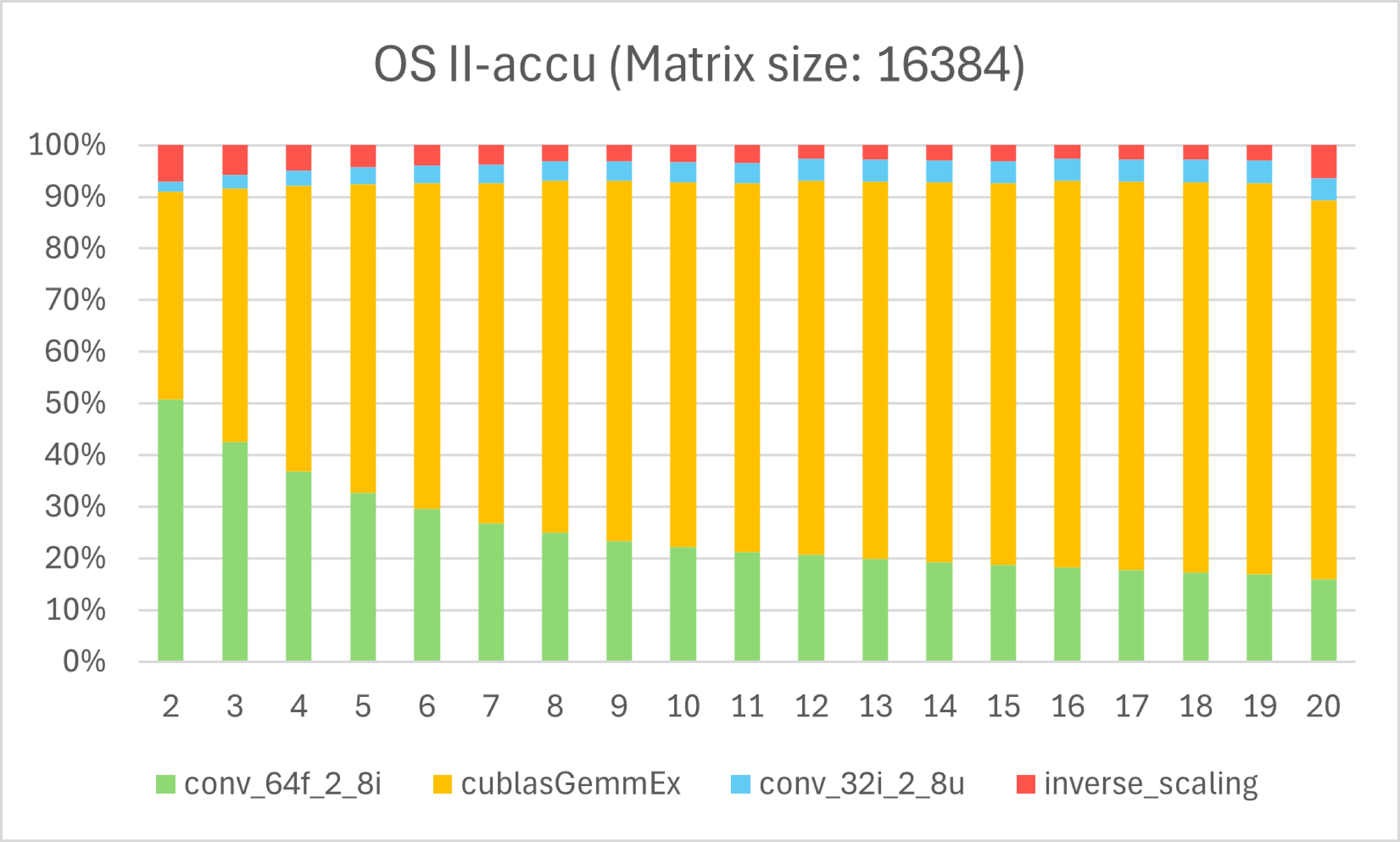}
\end{minipage}

\noindent
\begin{minipage}[b]{0.49\hsize}\centering
\includegraphics[width=\hsize]{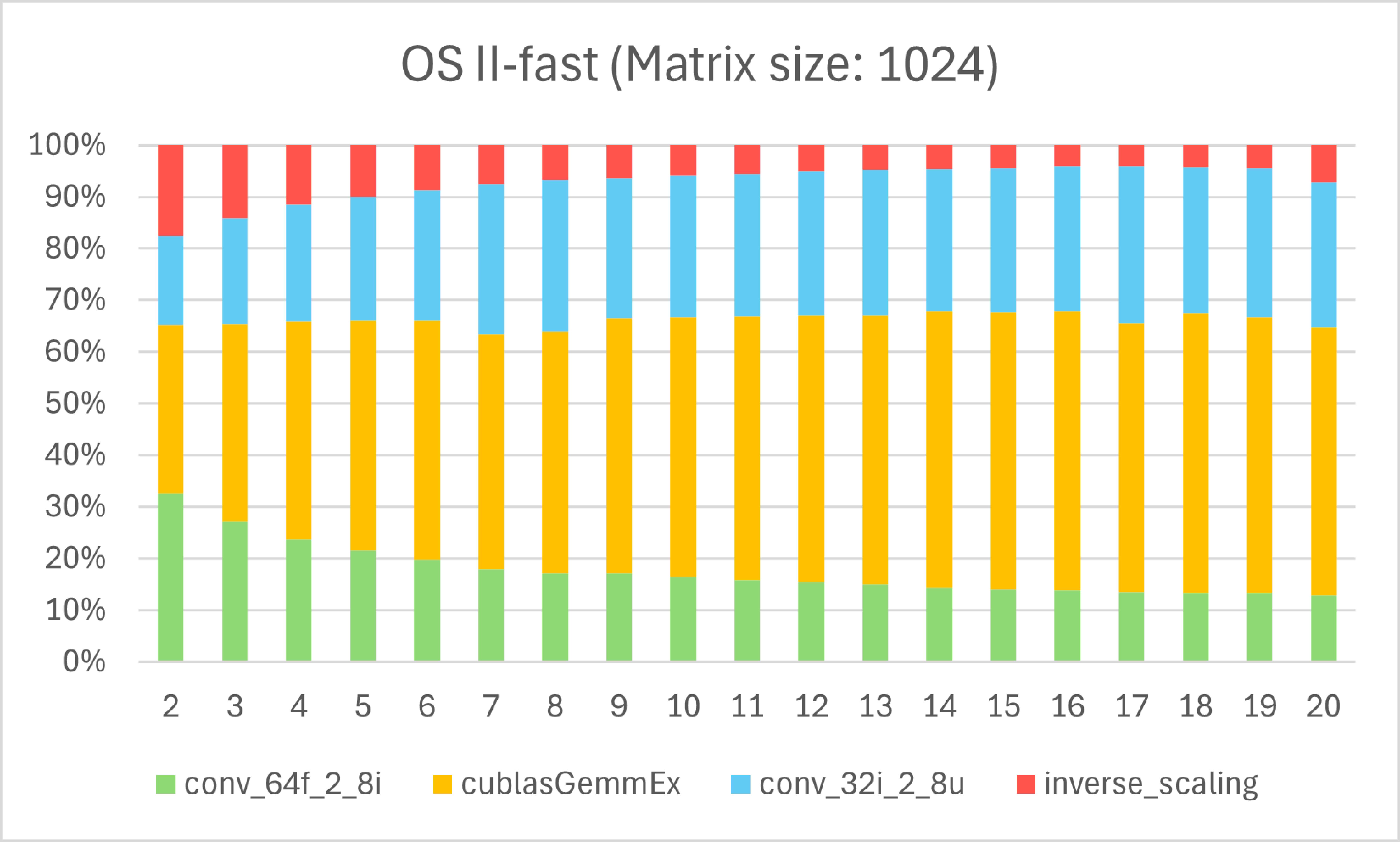}
\end{minipage}
\begin{minipage}[b]{0.49\hsize}\centering
\includegraphics[width=\hsize]{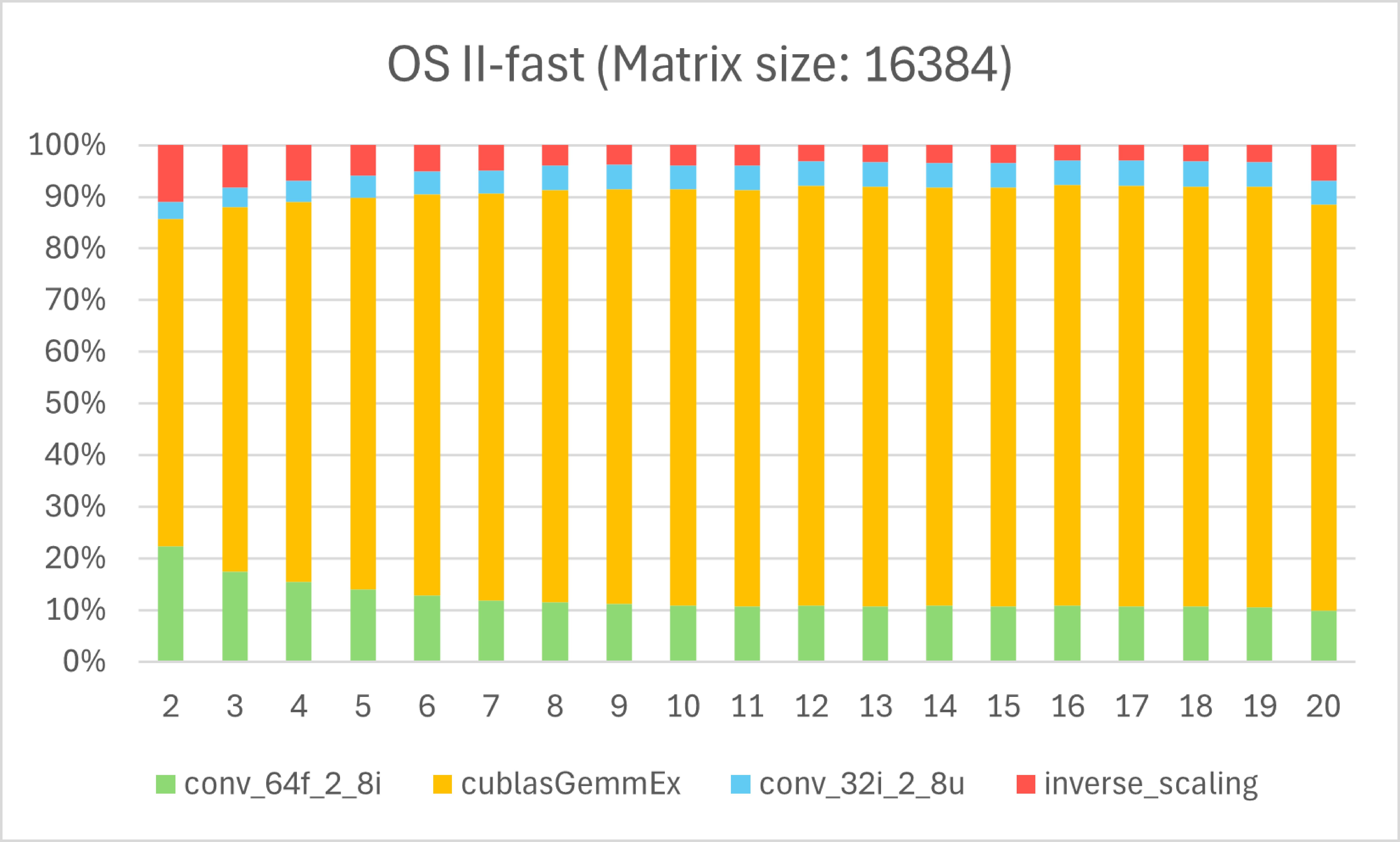}
\end{minipage}

\caption{Time breakdown for $p = r = q \in \{1024, 16384\}$ on NVIDIA GH200 Grace Hopper Superchip. Horizontal values represent number of moduli.}
\label{fig:int8-time-breakdown-gh200}
\end{figure*}

\begin{figure*}[htb]\centering
\noindent
\begin{minipage}[b]{0.49\hsize}\centering
\includegraphics[width=\hsize]{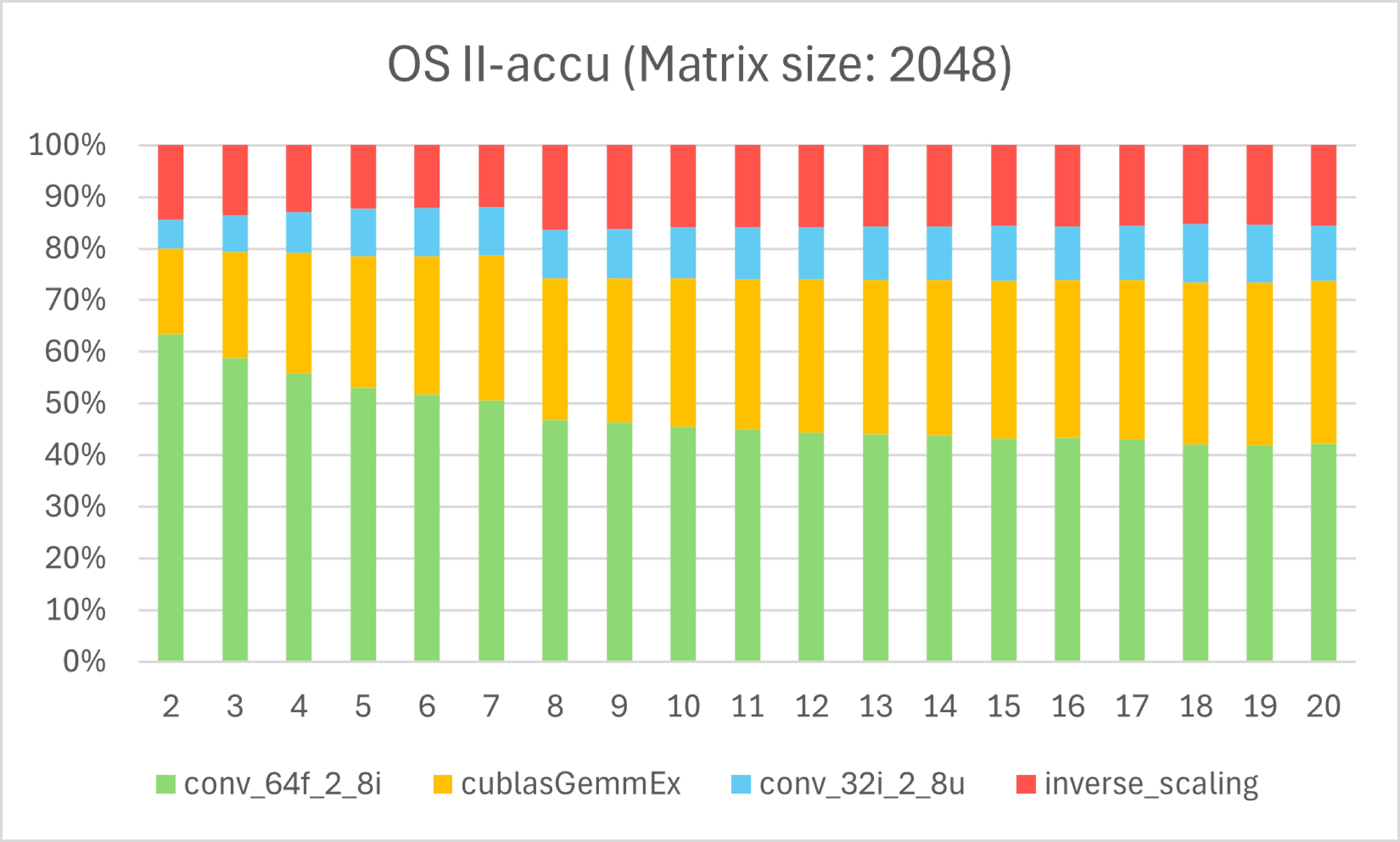}
\end{minipage}
\begin{minipage}[b]{0.49\hsize}\centering
\includegraphics[width=\hsize]{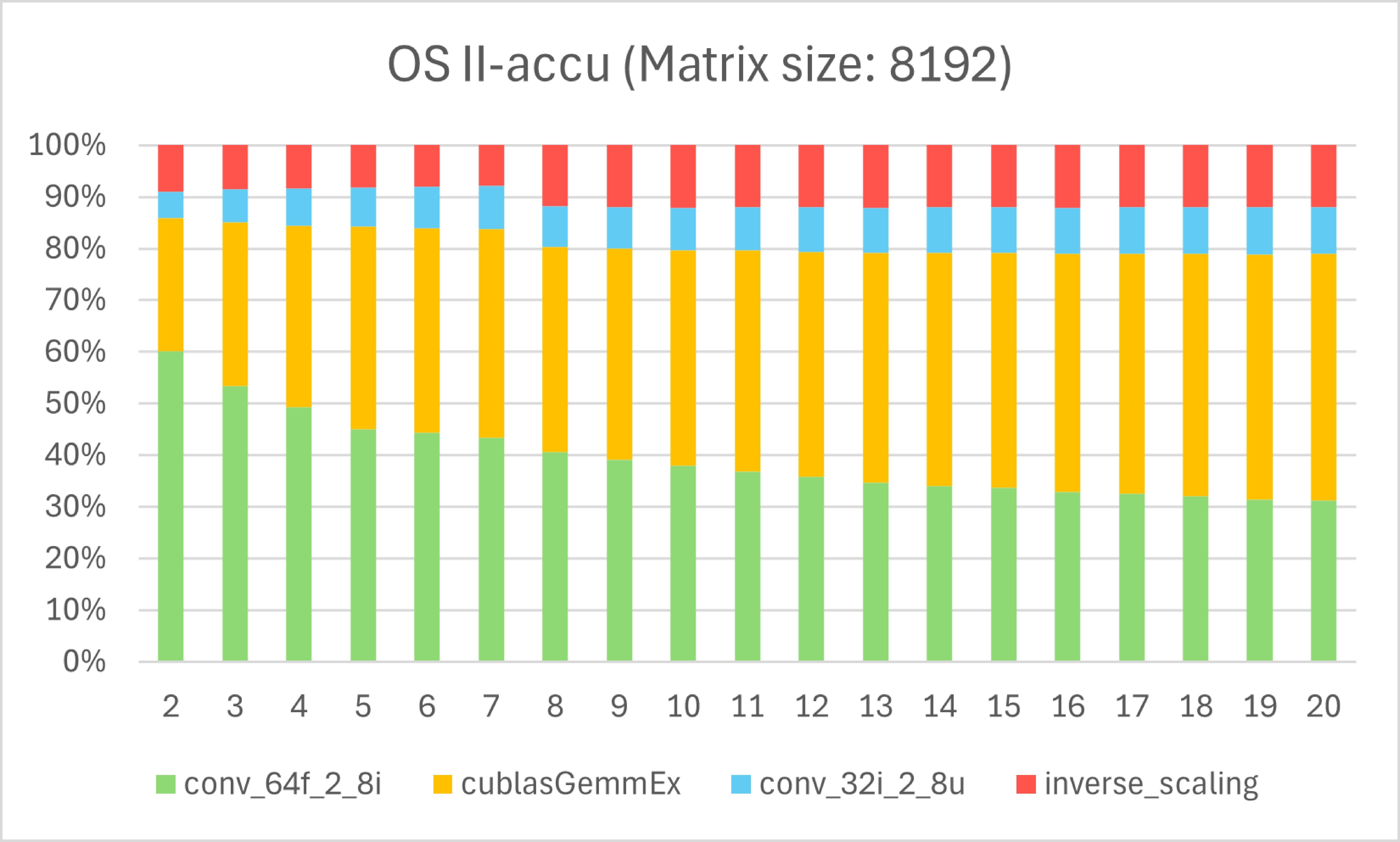}
\end{minipage}

\noindent
\begin{minipage}[b]{0.49\hsize}\centering
\includegraphics[width=\hsize]{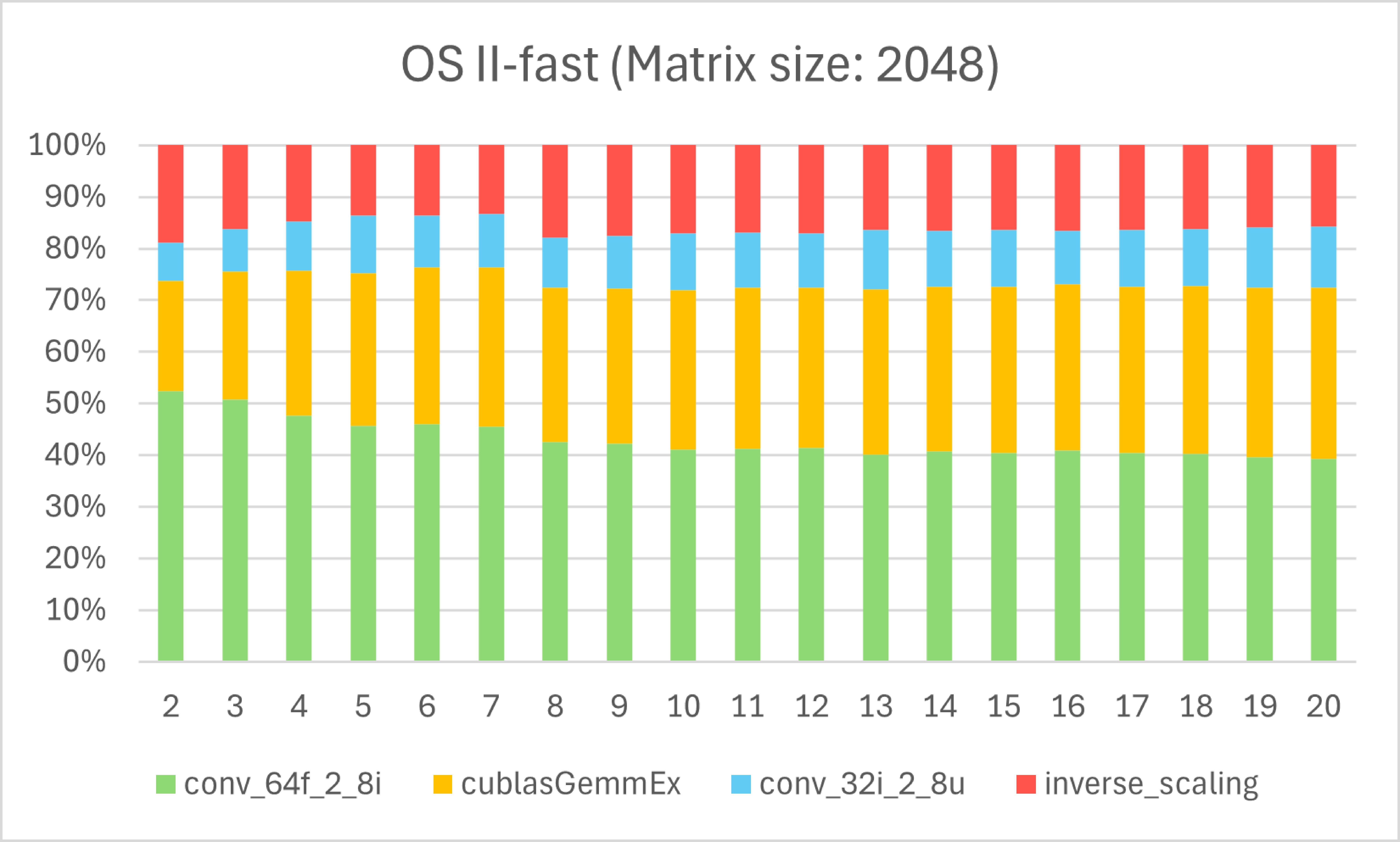}
\end{minipage}
\begin{minipage}[b]{0.49\hsize}\centering
\includegraphics[width=\hsize]{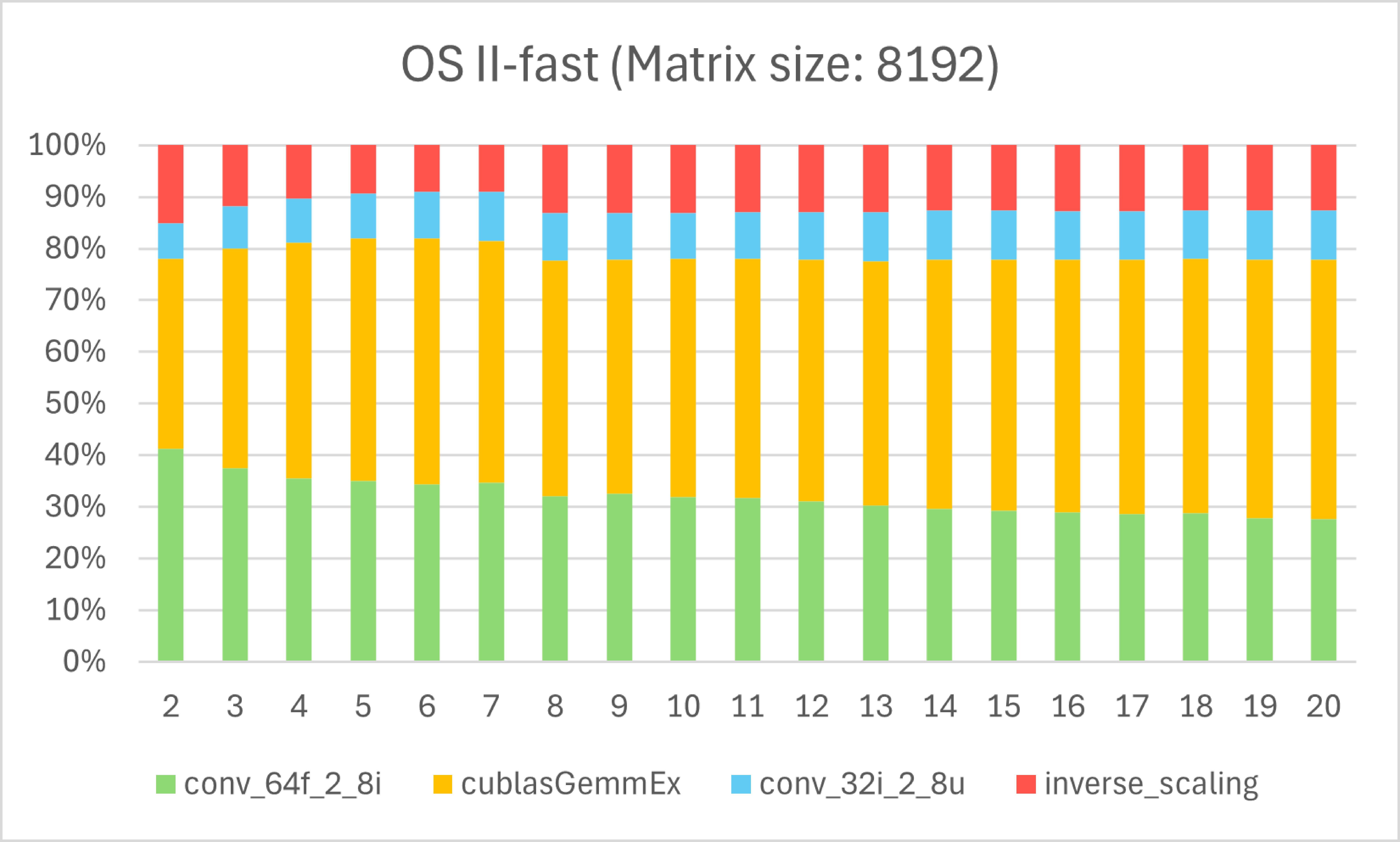}
\end{minipage}

\caption{Time breakdown for $p = r = q \in \{2048, 8192\}$ on NVIDIA GeForce RTX 4090 GPU. Horizontal values represent number of moduli.}
\label{fig:int8-time-breakdown-rtx}
\end{figure*}

Figures~\ref{fig:int8-time-breakdown-gh200} and \ref{fig:int8-time-breakdown-rtx} show the time breakdown of OS II-fast-$s$ and OS II-accu-$s$ for $s=2,3,\dots,20$.
For small problems, 
    the overhead of kernel launches becomes a performance bottleneck. 
For medium-sized problems, 
    while the kernel launch overhead is mitigated, 
    the conversion from double-precision matrices to INT8 matrices remains the main bottleneck, 
    and the adverse effect of the accumulation of matrix products, which correspond to conv\_32i\_2\_8u (line~\ref{alg:Oz2-int8-7} in Algorithm~\ref{alg:Oz2-int8}) and inverse\_scaling (lines~\ref{alg:Oz2-int8-8}--\ref{alg:Oz2-int8-10} in Algorithm~\ref{alg:Oz2-int8}), is also not negligible. 
For sufficiently large problems, 
    both the kernel launch overhead and the accumulation of matrix products become nearly negligible; 
    however, in OS II-accu-$s$, the conversion from double-precision matrices to INT8 matrices continues to be the dominant bottleneck because matrix multiplication is performed in the phase.

The peak performance of INT8 TCs on the GH200 is 1979~TOPS, while the peak performance of FP64 TCs is 67~TFLOPS. 
On GPUs such as the B200, the peak performance of INT8 TCs increases to 4500~TOPS, whereas the FP64 TCs achieve 40~TFLOPS. 
In such environments, where INT8 TCs performance improves while FP64 TCs performance decreases compared to the GH200, emulation techniques are expected to become increasingly important.

\subsection{Using FP64 for Quad-word arithmetic emulation}

The computational environment consisted of a PC with an Intel\textsuperscript{\textregistered} Core\textsuperscript{\texttrademark} i7-8665U processor (4 cores) and an Intel\textsuperscript{\textregistered} Core\textsuperscript{\texttrademark} i9-10980XE processor (18 cores), MATLAB 2024b, and Windows 10.
Hereafter, we simply refer to them as the Core i7 and Core i9, respectively.
The code was implemented as a MATLAB executable (MEX) and compiled with Visual Studio 2022. The compiler flags \texttt{/openmp} and \texttt{/arch:AVX2} were used for the Core i7, and \texttt{/openmp} and \texttt{/arch:AVX512} were used for the Core i9.

\begin{figure*}[htb]\centering
\noindent
\begin{minipage}[b]{0.33\hsize}\centering
\includegraphics[width=\hsize]{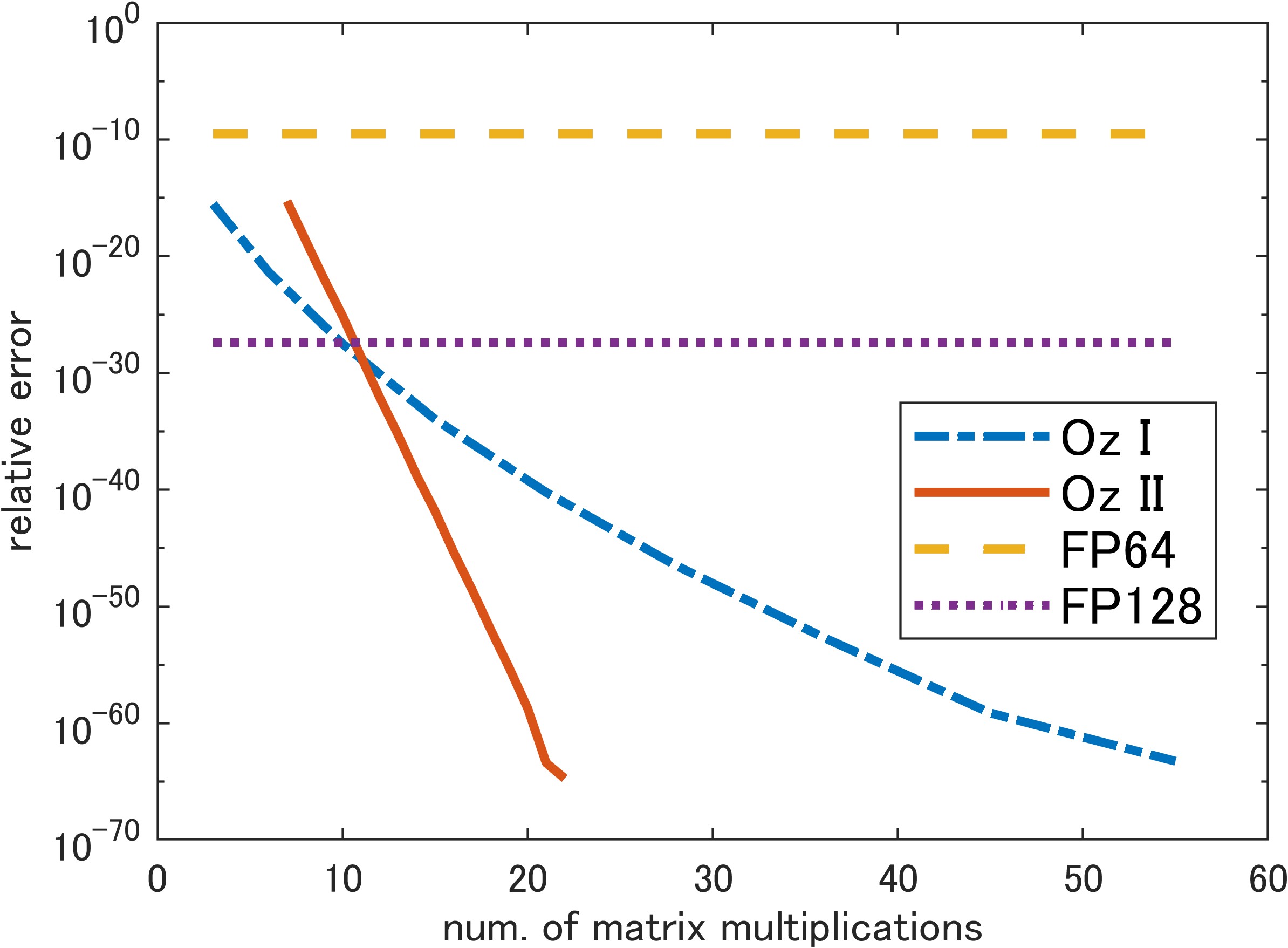}
\end{minipage}
\begin{minipage}[b]{0.33\hsize}\centering
\includegraphics[width=\hsize]{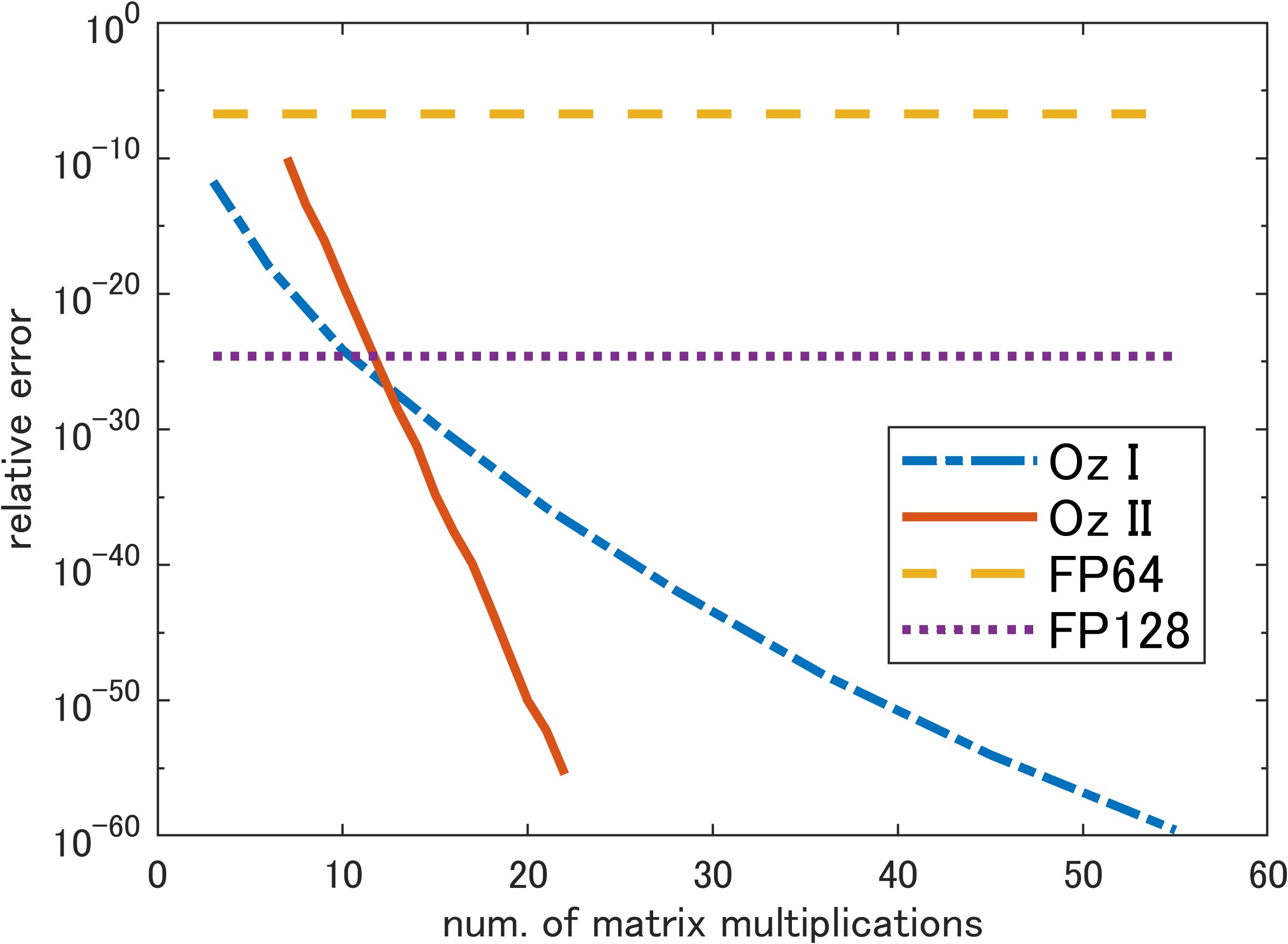}
\end{minipage}
\begin{minipage}[b]{0.33\hsize}\centering
\includegraphics[width=\hsize]{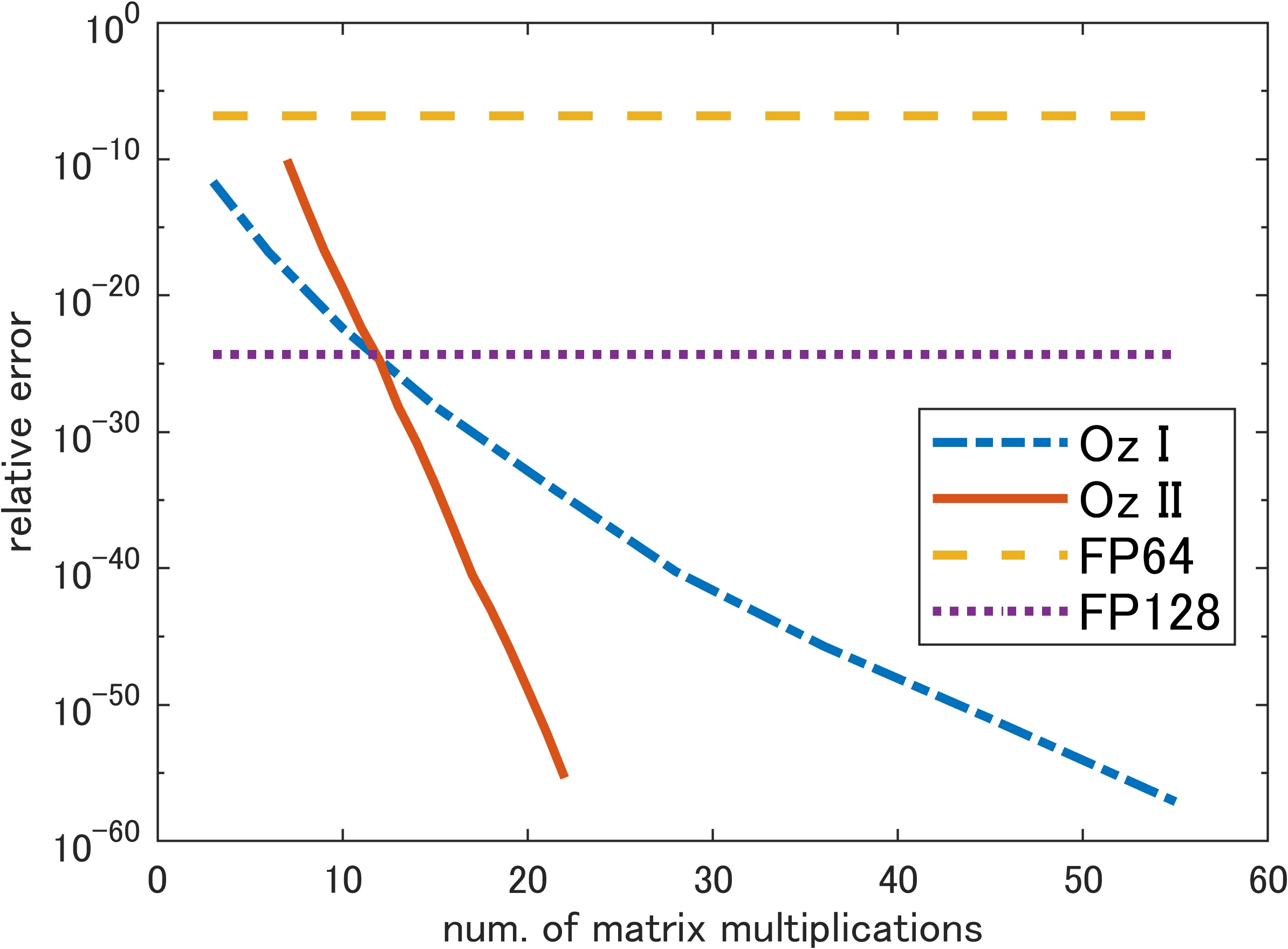}
\end{minipage}

\caption{Accuracy comparison for $p = r = q \in \{1000, 8000, 15000\}$ on Intel\textsuperscript{\textregistered} Core\textsuperscript{\texttrademark} i9-10980XE processor. Horizontal values represent number of moduli.}
\label{fig:FP64accuracy}
\end{figure*}

We set $v = 4$ in~\eqref{eq:multi-word}.
It indicates that
\begin{align*}
A & = A^{(1)} + A^{(2)} + A^{(3)} + A^{(4)}, \\
B & = B^{(1)} + B^{(2)} + B^{(3)} + B^{(4)}.
\end{align*}
$A^{(1)}$ and $B^{(1)}$ are generated by the $\mathtt{randn}$ function on MATLAB, and \eqref{eq:diff} are satisfied.
We use six-word arithmetic for Part 2-c and Part 3 in Ozaki scheme II.
Figure~\ref{fig:FP64accuracy} shows the maximum relative errors for Ozaki schemes~I and II for $p=q=r \in \{ 1000, 8000, 15000\}$ on the  Core-i9.
When the number of matrix multiplications is small, the accuracy of Ozaki scheme II is worse than that of Ozaki scheme~I. However, when the number of matrix multiplications increases, Ozaki scheme~II shows better accuracy than Ozaki scheme I. This is because the accuracy of Ozaki scheme~II improves linearly with respect to the number of matrix multiplications on a logarithmic scale.
The reason is that $\max_i m_i \approx \min_i m_i$ is satisfied, as explained in the previous section.

\begin{figure*}[htb]\centering
\noindent
\begin{minipage}[b]{0.49\hsize}\centering
\includegraphics[width=\hsize]{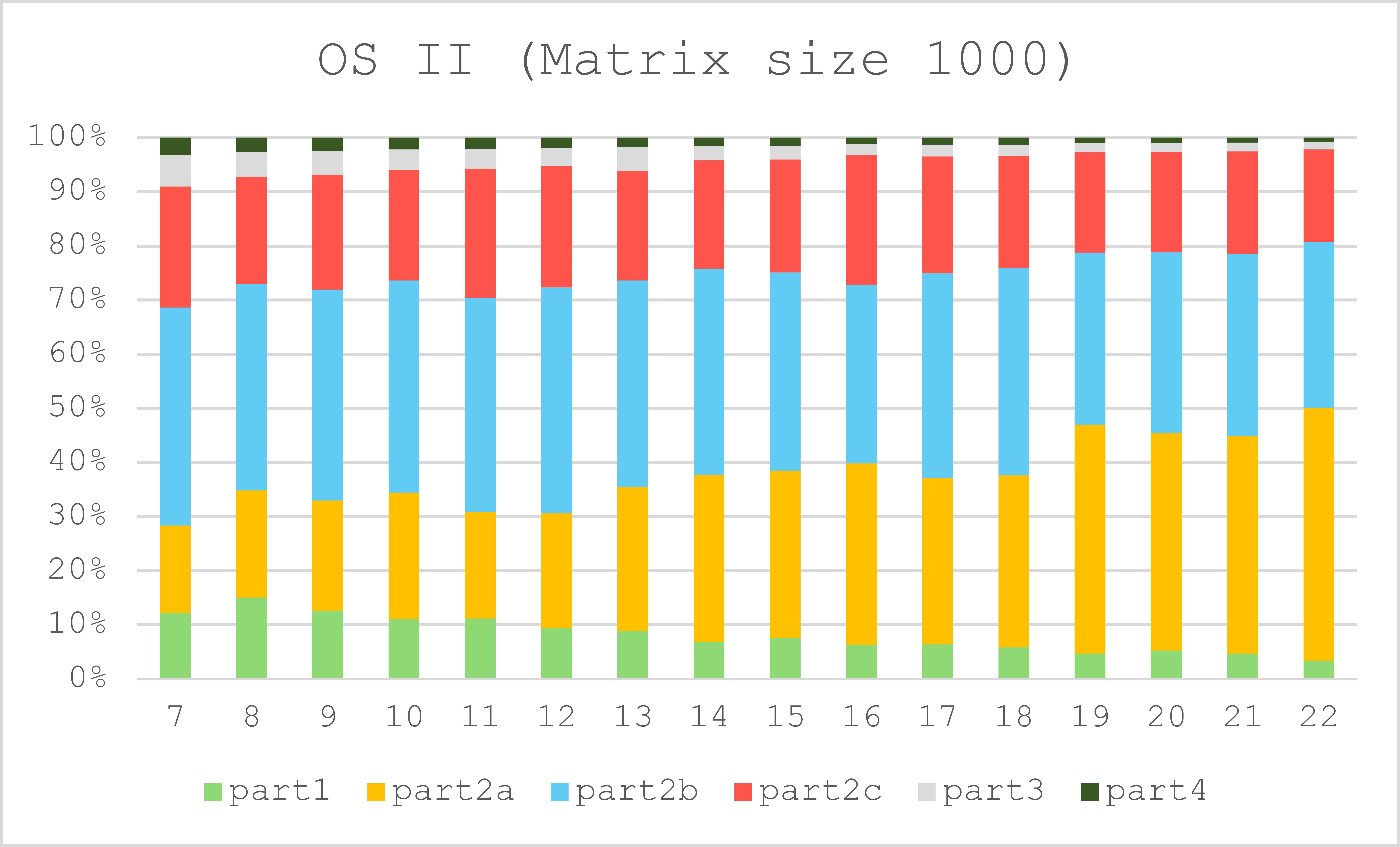}
\end{minipage}
\begin{minipage}[b]{0.49\hsize}\centering
\includegraphics[width=\hsize]{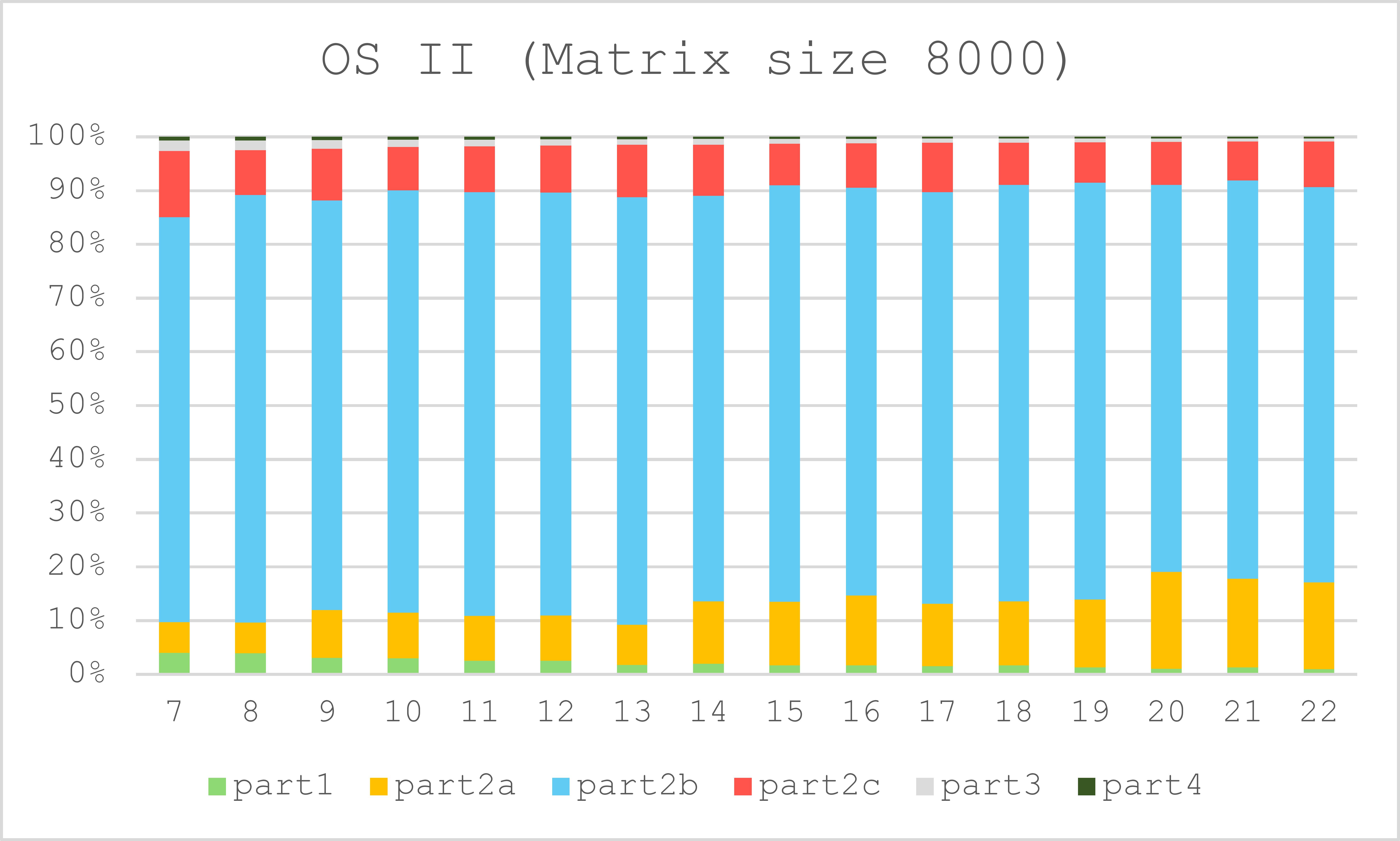}
\end{minipage}

\caption{Time breakdown for $p = r = q \in \{1000, 8000\}$ on 
Intel\textsuperscript{\textregistered} Core\texttrademark{} i7-8665U processor. Horizontal values represent number of moduli.}
\label{fig:FP64ratio_i7}
\end{figure*}

\begin{figure*}[htb]\centering
\noindent
\begin{minipage}[b]{0.49\hsize}\centering
\includegraphics[width=\hsize]{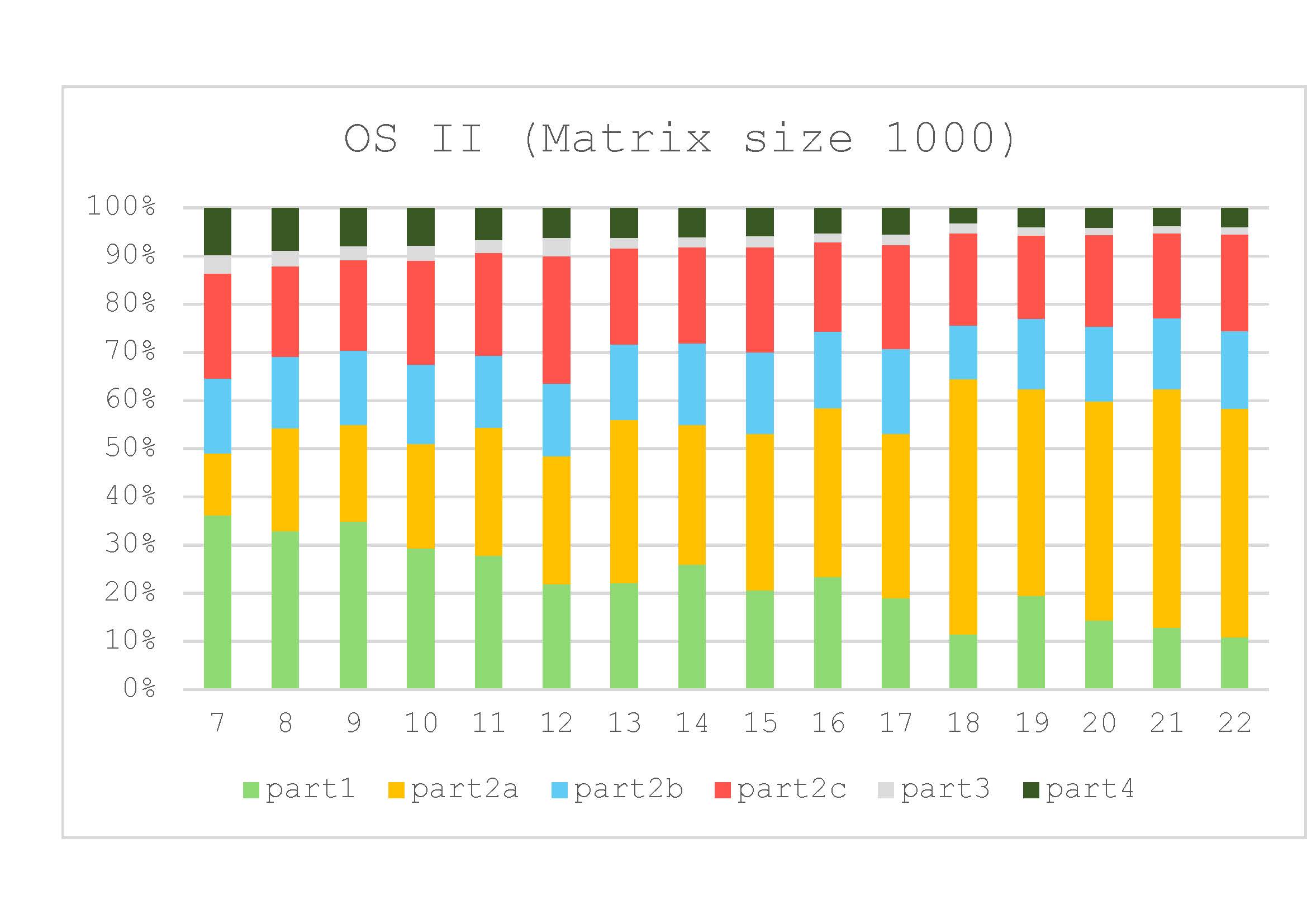}
\end{minipage}
\begin{minipage}[b]{0.49\hsize}\centering
\includegraphics[width=\hsize]{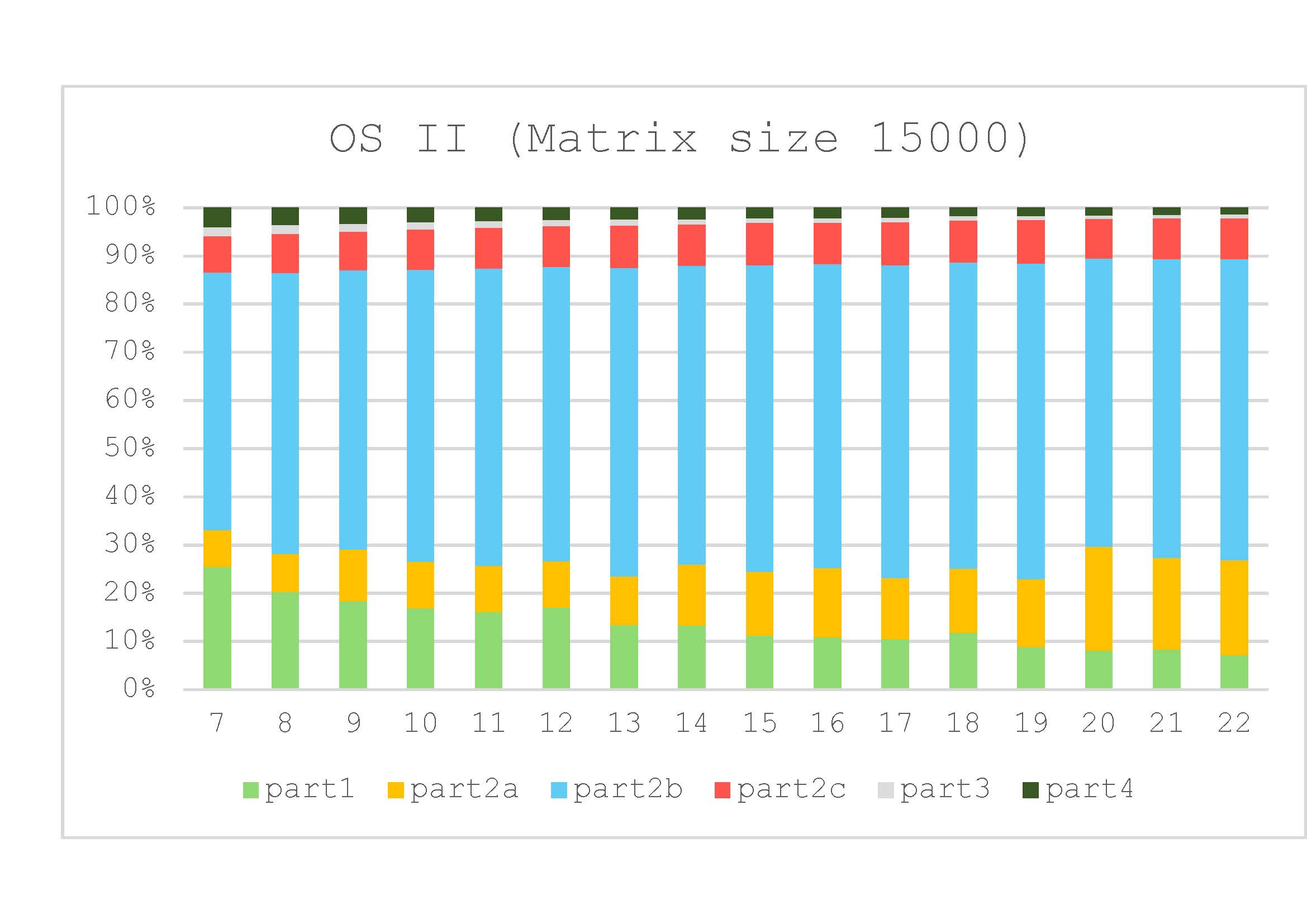}
\end{minipage}

\caption{Time breakdown for $p = r = q \in \{1000, 15000\}$ on Intel\textsuperscript{\textregistered} Core\textsuperscript{\texttrademark} i9-10980XE processor. Horizontal values represent number of moduli.}
\label{fig:FP64ratio_i9}
\end{figure*}

Figures~\ref{fig:FP64ratio_i7} and \ref{fig:FP64ratio_i9} show the proportion of computation time for each part across different matrix sizes. 
Although the computation time of Part 2-b is ideally expected to be dominant, it is evident that it does not account for the largest portion of the total computation time when $n=1000$.
As the matrix size increases, the proportion of time spent on matrix multiplication becomes higher, indicating that the method becomes more dependent on GEMM.

Tables~\ref{tab:fp64timei7} and \ref{tab:fp64timei9} show the throughput in GFLOPS when using the Core i7 and Core i9, respectively.
From Figure~\ref{fig:FP64accuracy},
Ozaki scheme I-10 (with 55 matrix multiplications) and Ozaki scheme II-22 exhibit comparable accuracy. 
However, as shown in Table 5, Ozaki scheme II is slower when $n=1000$, whereas it demonstrates superior performance for values of $n$ other than 1000.
Ozaki scheme~II achieved a speedup of up to 2.29x on the Core i7 at $n=4000$ and up to 2.12x on the Core i9 at $n=8000$.

\begin{table}[htb]
\centering
\caption{Throughput in GFLOPS for $p=q=r \in \{1000, 4000, 8000\}$ on Intel\textsuperscript{\textregistered} Core\texttrademark{} i7-8665U processor \label{tab:fp64timei7}}
\begin{tabular}{lccc}
\hline
    Methods $\backslash$ Matrix size & $1000$ & $4000$ & $8000$ \\
\hline
    DGEMM        & 99.5   & 105 & 114 \\
    OS I-$6$   & 2.32   & 4.35 & 4.32 \\
    OS I-$7$   & 2.32   & 3.33 & 3.02 \\
    OS I-$8$   & 2.32   & 2.63 & 2.31 \\
    OS I-$9$   & 1.96   & 2.11 & 1.88 \\
    OS I-$10$  & 1.60   & 1.75 & 1.51 \\
    OS II-$19$   & 1.35   & 3.45 & 4.10 \\
    OS II-$20$   & 1.31   & 2.99 & 3.58 \\
    OS II-$21$   & 1.18   & 2.89 & 3.45 \\
    OS II-$22$   & 1.20   & 2.81 & 3.47 \\
\hline
\end{tabular}
\end{table}

\begin{table}[htb]
\centering
\caption{Throughput in GFLOPS for $p=q=r \in \{1000, 8000, 15000\}$ on Intel\textsuperscript{\textregistered} Core\textsuperscript{\texttrademark} i9-10980XE processor \label{tab:fp64timei9}}
\begin{tabular}{lccc}
\hline
    Methods $\backslash$ Matrix size & $1000$ & $8000$ & $15000$ \\
\hline
    DGEMM        & 228   & 1073 & 1092    \\
    OS I-$6$   & 6.51   & 29.6 & 30.5   \\
    OS I-$7$   & 5.42   & 23.5 & 25.1   \\
    OS I-$8$   & 4.59   & 20.3 & 21.0   \\
    OS I-$9$   & 4.24   & 17.0 & 17.9   \\
    OS I-$10$  & 3.79   & 12.4 & 15.1   \\
    OS II-$19$   & 4.46   & 27.3 & 33.7   \\
    OS II-$20$   & 4.54   & 26.2 & 30.2   \\
    OS II-$21$   & 4.19   & 26.6 & 29.8   \\
    OS II-$22$   & 4.31   & 26.3 & 28.8   \\
\hline
\end{tabular}
\end{table}

\section{Conclusion}
\label{sec:Conclusion}

In this study, we developed a novel algorithm, referred to as Ozaki scheme~II, for emulating matrix multiplication. 
Furthermore, the effectiveness of multi-word arithmetic was also demonstrated. 
Future challenges include further optimization of the implementation and a detailed analysis of the rounding errors introduced by the proposed method, which will be addressed in future work.
We also plan to conduct further experiments on GPUs with higher INT8 Tensor Core performance, such as the B200, to explore the potential of our method on next-generation architectures.

\section*{Appendix: Related Work}
\label{sec:Appendix}

After submitting this paper, several related methods and applications have been studied. These include extensions of GPU-based Ozaki Scheme I~\cite{Angelika}, implementations and energy-efficiency evaluations of GPU-based Ozaki Scheme II~\cite{10.1145/3731599.3767539}, extensions of Ozaki Scheme II to complex matrices~\cite{uchino2025emulationcomplexmatrixmultiplication}, rounding error analyses for Ozaki Scheme II~\cite{uchino2026erroranalysismatrixmultiplication}, FP8-TC-oriented variants of Ozaki Scheme II~\cite{uchino2026doubleprecisionmatrixmultiplicationemulation}, and applications in cuEST, a quantum chemistry computation library~\cite{nvidia_cuest_emulation}.



\begin{dci}
The authors declare no conflict of interest.
\end{dci}

\begin{funding}
This study was partially supported by the JSPS Grant-in-Aid for Research Activity Start-up No. 24K23874, and the JSPS KAKENHI Grant Nos. 23K28100, 25H01109, and 25K03126.
\end{funding}

\begin{sm}
 Not applicable.
\end{sm}

\bibliographystyle{SageH}


\begin{biogs}
Katsuhisa Ozaki is a full professor in the Department of Mathematical Sciences at the Shibaura Institute of Technology. He received his Ph.D. in engineering from Waseda University in 2007. He was an Assistant Professor (2007-2008) and a Visiting Lecturer (2008-2009) at Waseda University. At Shibaura Institute of Technology, he has served as Assistant Professor (2010-2013) and Associate Professor (2013-2019) and currently is a Professor since 2019. His research interests include reliable computing, particularly addressing rounding error problems in finite-precision arithmetic. He mainly focuses on numerical linear algebra and develops fast and accurate algorithms.

Yuki Uchino is a postdoctoral researcher at RIKEN R-CCS. 
He received his Ph.D. in engineering from Shibaura Institute of Technology in 2024. 
His research interests include reliable computing, numerical linear algebra, and highly accurate algorithms. 

Toshiyuki Imamura is a team principal of the Large-scale Parallel Numerical Computing Technology Team at RIKEN R-CCS, and is responsible for developing numerical libraries on Fugaku. 
He received his Diploma and Doctorate in Applied Systems and Sciences from Kyoto University in 1993 and 2000. 
He was a Researcher at CCSE, JAERI (1996-2003), a visiting scientist at HLRS (2002), and an associate professor at the University of Electro-Communications (2003-2012). 
His research interests include HPC, auto-tuning technology, and parallel eigenvalue computation. 
His research group won the HPL-MxP ranking (2020-2021) and was nominated as the Gordon Bell Prize finalist in SC05, SC06, and SC20.
\end{biogs}

\end{document}